\documentclass[prd,amsmath,aps,floats,amssymb, floatfix, superscriptaddress,
twocolumn, nofootinbib]{revtex4-1}

\usepackage{tabularx}
\usepackage{amsmath,amssymb,natbib,latexsym,times}

\usepackage{booktabs}

\usepackage{graphicx}
\usepackage{dcolumn}
\usepackage{bm}
\usepackage{hyperref}
\usepackage{physics}
\usepackage{comment}
\usepackage{multirow}
\usepackage{url}
\usepackage{mathrsfs}

\usepackage[dvipsnames]{xcolor}
\usepackage{orcidlink}

\usepackage{graphicx}

\newcommand{\mnras}{Monthly Notices of the Royal Astronomical Society}

\newcommand{\jcap}{Journal of Cosmology and Astroparticle Physics}

\renewcommand{\apj}{The Astrophysical Journal}

\newcommand{\apjs}{The Astrophysical Journal Supplement Series}
\newcommand{\aap}{Astronomy and Astrophysics}
\newcommand{\pasj}{Publ.~Astron.~Soc.~Japan}

\newcommand{\bk}{{\bf k}}

\newcommand{\KCDM}{\mbox{$\Omega_K$-$\Lambda$CDM}}
\newcommand{\fCDM}{\mbox{$f\Lambda$CDM}}
\newcommand{\LCDM}{\mbox{$\Lambda$CDM}}
\newcommand{\deltab}{\delta_{\rm b}}
\newcommand{\parfrac}[2]{\frac{\partial #1}{\partial #2}}

\begin{document}

\title[]{Flat to nonflat: \\ Calculating nonlinear power spectra
of biased tracers  for nonflat \LCDM~model}

\author{Ryo~Terasawa\orcidlink{0000-0002-1193-623X}}
\email{ryo.terasawa@ipmu.jp}
\affiliation{Kavli Institute for the Physics and Mathematics of the Universe (WPI), The University of Tokyo Institutes for Advanced Study (UTIAS), The University of Tokyo, Chiba 277-8583, Japan}
\affiliation{Department of Physics, The University of Tokyo, Bunkyo, Tokyo 113-0031, Japan}
\affiliation{
Center for Data-Driven Discovery (CD3), Kavli IPMU (WPI), UTIAS, The University of Tokyo, Kashiwa, Chiba 277-8583, Japan
}
\author{Ryuichi~Takahashi\orcidlink{0000-0001-6021-0147}}
\affiliation{
Faculty of Science and Technology, Hirosaki University, 3 Bunkyo-cho, Hirosaki, Aomori 036-8561, Japan
}
\author{Takahiro~Nishimichi\orcidlink{0000-0002-9664-0760}}
\affiliation{Department of Astrophysics and Atmospheric Sciences, Faculty of Science, Kyoto Sangyo University, Motoyama, Kamigamo, Kita-ku, Kyoto 603-8555, Japan}
\affiliation{Center for Gravitational Physics and Quantum Information, Yukawa Institute for Theoretical Physics, Kyoto University, Kyoto 606-8502, Japan}
\affiliation{Kavli Institute for the Physics and Mathematics of the Universe
(WPI), The University of Tokyo Institutes for Advanced Study (UTIAS),
The University of Tokyo, Chiba 277-8583, Japan}
\author{Masahiro~Takada\orcidlink{0000-0002-5578-6472}}
\affiliation{Kavli Institute for the Physics and Mathematics of the Universe (WPI), The University of Tokyo Institutes for Advanced Study (UTIAS), The University of Tokyo, Chiba 277-8583, Japan}
\affiliation{
Center for Data-Driven Discovery (CD3), Kavli IPMU (WPI), UTIAS, The University of Tokyo, Kashiwa, Chiba 277-8583, Japan
}

\begin{abstract}
 The growth of 
 large-scale structure, together with 
 the geometrical information of cosmic expansion history and cosmological distances, can be used to obtain
 constraints on the spatial curvature of the universe that probes the early universe physics, whereas modeling the nonlinear growth in a nonflat universe is still challenging due to 
 computational expense of simulations in a high-dimensional cosmological parameter space.
In this paper, we develop an approximate method to compute the halo-matter and halo-auto power spectra
for nonflat \LCDM~model, from quantities representing the nonlinear evolution of the corresponding flat \LCDM~model, based on the separate universe (SU) method. 
By utilizing the fact that the growth response to long-wavelength fluctuations (equivalently the curvature), $T_{\deltab}(k)$, is approximated by the response to the Hubble parameter, $T_h(k)$, our method allows one to estimate the nonlinear power spectra in a nonflat universe efficiently from the power spectra of the flat universe. We use $N$-body simulations to show that the estimator can provide the halo-matter (halo-auto) power spectrum at $\sim 1\%$ ($\sim 2\%$ ) accuracy up to $k \simeq 3 (1) \, h {\rm Mpc}^{-1}$ even for a model with
large curvature $\Omega_K = \pm 0.1$.
Using the estimator we can extend the prediction of the existing emulators such as  {\tt Dark Emulator} to nonflat models without degrading their accuracy.
Since the response to long-wavelength fluctuations is also a key quantity for estimating the super sample covariance (SSC), we discuss that the approximate identity $T_{\deltab}(k) \approx T_h(k)$ can be used
to calculate the SSC terms analytically. 
\end{abstract}

\maketitle

\section{Introduction}

The spatial curvature of the universe ($\Omega_K$) is an important quantity that characterizes the geometry of the universe and probes the physics of the early universe including the mechanism of inflation. 
The curvature of the universe influences various observables, including the spatial distribution of galaxies, through both its geometric properties and its impact on the growth of large-scale structures (LSS). The geometrical constraint, inferred from 
the primary CMB anisotropy information of the {\it Planck} data~\citep{planck_collaboration_2020},
is given as
$\Omega_K=-0.044^{+0.018}_{-0.015}$ (68\% CL, {\it Planck} TT, TE, EE+lowE), implying a $2\sigma$ hint of the close geometry, although most of the constraints are consistent with flat universe \citep[e.g.][]{2017MNRAS.470.2617A,2021PhRvD.103f3511K, 2021ApJ...908...84V,2021MNRAS.506L...1D,  2021PhRvD.103h3533A, 2021PDU....3300851V, 2022ApJ...939...37L, 2022arXiv220705766D, 2023arXiv230910034T}.
This indicates the importance of constraining the curvature from LSS as an independent probe.
Furthermore, to obtain more precise constraints on the curvature close to the amplitudes of primordial fluctuations, $|\Omega_K| \sim 10^{-4}$, it is crucial to combine observations of these two effects~\citep[e.g.][]{curvature_endgame}.
However, achieving this requires development of a theoretical model that adequately incorporates mode coupling effects. Mode coupling refers to the phenomenon where the non-linear nature of gravity causes fluctuations of different wavelengths to interact with each other. 
Constructing such a model is generally challenging. Despite its inherent difficulty, incorporating small-scale information is vital for obtaining stronger constraints on cosmological parameters.
For flat cosmological models in particular, fitting formulae or emulators based on the predictions of $N$-body simulations have been developed~\citep[e.g.][]{1996MNRAS.280L..19P,Smith03,Coyote1,Coyote2,Takahashi12,MiraTitan1,MiraTitan2,Nishimichi_2019,2020MNRAS.492.5226R,2021MNRAS.502.1401M,2019MNRAS.486.1448S,2021MNRAS.505.2840E}. 
For example, {\tt Dark Emulator}~\citep{Nishimichi_2019} predicts the basic statistical quantities of dark matter halos (halos in what follows for simplicity)
such as their abundance as a function of mass and the halo power spectrum down to the nonlinear scale. With these nonlinear predictions of the halo statistics, together with the halo occupation distribution (hereafter, HOD) description, Ref.~\cite{HSCY3_Miyatake} analyzed the two-point correlation function of galaxy clustering, galaxy-galaxy lensing and cosmic shear (so-called $3 \times 2$ pt analysis), and obtained tighter 
constraints on the cosmological parameters compared to the traditional analysis based on the linear bias 
model~\citep{HSCY3_Sugiyama}.

Theoretical models for the power spectrum in a nonflat universe are still in the development stage compared to models in flat universe due to the computational expense of covering a wider range in a multi-dimensional cosmological parameter space.
Because of this limitation of knowledge on 
nonlinear scales, in the $3 \times 2$ pt analysis of the DES Y3~\citep{2022arXiv220705766D} data they used only the linear scales. In this work, we provide a method to model the 
power spectrum down to nonlinear scales so that one can gain information on the curvature from small scales.

The effect of curvature on structure formation can be identified with the very long-wavelength density fluctuations in the \LCDM~model based on the separate universe (SU) approach
\cite{2011JCAP...10..031B,2013PhRvD..87l3504T,2014PhRvD..89h3519L,2014PhRvD..90j3530L,Wagner+15a}.
In~\cite{Terasawa+22}, we developed a method to compute the nonlinear matter power spectrum for nonflat cosmologies utilizing the SU approach.
We utilized the approximate identity that states the effect of these long-wavelength density fluctuations on structure formation is well reproduced by the response to the Hubble parameter $h$~\citep{2014PhRvD..90l3523S,Terasawa+22}.
Using the response to long-wavelength fluctuations modeled in this way, we have shown that the model of the matter power spectrum for flat universes,whose accurate model calculation is already available from the fitting formula or emulator, can be extended to calculate the nonlinear matter power spectrum for a universe with non-zero curvature.

In this work, we apply the SU approach to predict the halo-matter and halo-auto power spectra down to the nonlinear scales for nonflat cosmologies.
We will verify that the approximate identity holds for the halo-matter and halo-auto power spectrum responses and construct the estimators of these power spectra using the response to the Hubble parameter. The nonlinear prediction of these spectra
together with the HOD prescription provides a way to predict the galaxy-galaxy lensing or galaxy clustering data down to small scales~\citep[e.g.][]{HSCY1_Miyatake,HSCY3_Miyatake}.

Another application of the modeling of the response to long-wavelength fluctuations is to compute
the super sample covariance (SSC) \citep{2013PhRvD..87l3504T}, which is the sample variance contribution
caused by mode coupling with long-wavelength fluctuations.
SSC for the power spectrum can be computed using its response to the super survey modes~\citep[e.g.][]{2014PhRvD..90l3523S}.
We also provide a way to compute the response to the super survey modes using the power spectrum response to $h$, which enables us to utilize a fitting formula or an emulator validated only for flat geometry.

In the literature, there are analytical methods to compute the response, with the perturbation theory~\citep[e.g.][]{2016JCAP...09..007B} or halo model~\citep[e.g.][]{cosmolike}. 
Using these analytical methods, we can calculate the response quickly. 
However, these models suffer from the limitation of scales or inaccuracy.
On the other hand, measuring the SSC from 
the scatter among an ensemble of simulations~\citep[e.g.][]{2023PhRvD.108d3521B} is expected to provide an accurate covariance matrix with a relatively high computational cost.
The size of the data vector is expected to become larger for future surveys, especially the ones using multi-tracer, for which the number of simulations needed could be $\mathcal{O}(10^{2})-\mathcal{O}(10^{3})$.
In this paper, our model is compared with simulations, and shown to be more accurate than the analytical methods used in the literature, while the computational cost is much cheaper than a simulation ensemble.
Our method for computing the total response would be implemented into the Core Cosmology Library~\citep{CCL}.

This paper is organized as follows. In Sec.~\ref{sec:SUapproach} we first review the SU approach for the nonlinear matter power spectrum in Ref.~\cite{Terasawa+22} and generalize it to predict the halo-matter or halo power spectrum. In Sec.~\ref{sec:SSC} we introduce our method to compute the total response to the super survey modes using the response to $h$. In Sec.~\ref{sec:simulation} we describe details of $N$-body simulations for flat and nonflat \LCDM~ models used in this paper. In Sec.~\ref{sec:results} we present the main results of this paper and show a numerical validation of our methods. 
Sec.~\ref{sec:conclusion} is devoted to discussion and conclusion. 
In Appendix~\ref{sec:derivation} we describe the details of how to compute the power spectrum response to the long-wavelength fluctuations.
Throughout the paper, we assume \LCDM~model and flat geometry for the fiducial cosmology. We also assume that the SSC is calculated in this fiducial cosmology.

\section{SU approach for $P(k;\Omega_K)$}\label{sec:SUapproach}

\subsection{SU approach for  $P_{\rm mm}(k;\Omega_K)$}

In this section we briefly review
the SU approach to predict the nonlinear matter power spectrum $P_{\rm mm}(k)$ for 
a nonflat universe following Ref.~\cite{Terasawa+22}.

First, since the effect of the curvature or the long-wavelength density fluctuations on structure formation appears only in the late universe, throughout this paper we consider a model where structure formation in the early universe is identical to that of the fiducial flat universe. Specifically, we keep the parameters
\begin{align}
\left\{\omega_{\rm c}, \omega_{\rm b}, A_s, n_s\right\},
\label{eq:fixed_cosmological_parameters}
\end{align}
fixed, where $\omega_{\rm c}(\equiv \Omega_{\rm c}h^2)$ and $\omega_{\rm b}(\equiv \Omega_{\rm b}h^2)$ are the physical density parameters of CDM and baryon, respectively, and
$A_s$ and $n_s$ are the amplitude (at the pivot scale $k_{\rm pivot}=0.05\,{\rm Mpc}^{-1}$)
and the spectral tilt of the power spectrum of primordial curvature perturbations.
Note that we fix
the sum of neutrino masses
so that the early universe physics remains unchanged, and treat its impact only through the linear transfer function of 
total matter fluctuations at $z=0$ \citep[see Refs.][for more details]{Nishimichi_2019,2021arXiv210804215B}.
These models with the four fixed parameters (Eq.~\ref{eq:fixed_cosmological_parameters}) share the same 
linear power spectrum $P_L(k)$ 
at sufficiently high redshifts $z_i \gg 1$.
The remaining parameters that affect
structure formation
are $\Omega_K$ and $h$ within the nonflat \LCDM\, model. Note that the density parameter for the cosmological constant, $\Omega_\Lambda$, is automatically determined once the parameters above are all fixed.

In the SU approach, the effects of the background density modulation $\deltab(t)$ in a flat universe are interpreted as the local effective cosmology with modified background density 
\begin{align}
    \bar{\rho}_{\rm m}(t) = \bar{\rho}_{{\rm m}f}(t) \left[1 + \deltab(t)\right],
\end{align}
and non-zero curvature corresponding to $\deltab$ (see below).
Hence it gives a mapping between nonflat 
and flat \LCDM~models. Hereafter we call the two models
\KCDM~and \fCDM, respectively, and we denote quantities in the \fCDM~model by sub/superscript ``$f$''. We assume $\delta_{\rm b}(t)$ evolves according to the linear growth factor $D_f(t)$
as $\delta_{\rm b}(t)\propto D_f(t)$.

The curvature and overdensity are related to each other via
\begin{align}
    \Omega_{K} = -\frac{5\Omega_{\mathrm{m}}}{3}\frac{{\delta_{\rm b}(t)}}{D_f(t)}.
\end{align}
The remaining parameter, $h$, is mapped as
\begin{align}
    h_f = h (1 - \Omega_K)^{1/2}.
\end{align}
Note that the redshifts in these two cosmologies
at a given cosmic time $t$ are related as
\begin{align}
    (1+z_f)[1+\deltab(z_f)]^{1/3} = 1+z.
    \label{eq:scale_factor_mapping}
\end{align}

Inversely, we can treat the nonflat universe as an overdense/underdense region in the corresponding flat universe, and the power spectrum in the target nonflat universe can be approximated by Taylor expansion around the flat universe as
discussed in Ref.~\cite{Terasawa+22}:
\begin{align}
{P}(k,z;\Omega_K)&\simeq P^f(k,z_f;\delta_{\rm b})\nonumber\\
&\hspace{-3em}\simeq \left.P^f(k,z_f)\right|_{\delta_{\rm b}=0}
\left[1+\left.\frac{\partial \ln P^f(k,z_f;\delta_{\rm b})}
{\partial \delta_{\rm b}}\right|_{{\rm G},\delta_{\rm b}=0}\delta_{\rm b}\right] \nonumber \\
&\hspace{-3em}\equiv \tilde{P}(k,z;\Omega_K),
\label{eq:ps_estimator}
\end{align}
where $\delta_{\rm b}\equiv \delta_{\rm b}(z_f)$. 
In the last equality on the r.h.s.,
we have put the tilde symbol $\tilde{\hspace{1em}}$ on $\tilde{P}(k,z;\Omega_K)$ to explicitly denote 
that $\tilde{P}$ is an ``estimator'' of the nonlinear 
matter power spectrum for the $\KCDM$ model.
We introduced subscript ``G" to $\partial P^f(k)/\partial \delta_{\rm b}$ to emphasize that it is the growth response~\citep{2014PhRvD..89h3519L}.
The growth response can be computed as the difference between the power spectra at a fixed comoving wavenumber $k$ in the two SU cosmologies, omitting the dilation effect that originates from the difference in the scale factors (Sec.~\ref{sec:SSC}). Throughout the paper, the wavenumber $k$ refers to a given comoving wavenumber even if the corresponding physical scales differ among the cosmologies.
Note that the expansion of Eq.~({\ref{eq:ps_estimator}}) is applicable to not only the matter power spectrum but also the halo-matter and halo-auto power spectra as we will show below.

Furthermore, we define a normalized growth response $T_{\deltab}^{\rm mm}(k)$ as 
\begin{align}
T_{\delta_{\rm b}}^{\rm mm}(k)\equiv \left[2\frac{\partial \ln D(\deltab)}{\partial \delta_{\rm b}}\right]^{-1}
\left.\frac{\partial \ln P_{\rm mm}(k;\delta_{\rm b})}{\partial \delta_{\rm b}}\right|_{{\rm G},\delta_{\rm b}=0}.
\label{eq:T^mm_db}
\end{align}
The normalized response has an asymptotic behavior of $T_{\delta_{\rm b}}^{\rm mm}\rightarrow 1$ at the linear limit 
$k\rightarrow 0$, because $P_{\rm mm}(k,z)\propto D(z)^2P_L(k,z_i)$ in such linear regime.
The linear limit of the matter power spectrum growth response is given in 
Refs.~\cite{2011JCAP...10..031B,2013PhRvD..87l3504T} as
\begin{align}
2\frac{\partial \ln D(\deltab)}{\partial \delta_{\rm b}}
\approx \frac{26}{21}
.
\label{eq:dlnDdb}
\end{align}
Since we have a prediction for the linear power spectrum for {the} \KCDM~model, we only need to expand the nonlinear correction $B(k,z)$ defined as $P(k,z) \equiv P_L(k,z) B(k,z)$.
Using the normalized response $T_{\deltab}(k)$, we can write the expansion as
\begin{widetext}
\begin{align}
    \tilde{P}(k;\Omega_K) =  \left(\frac{D(z)}{D_f(z_f)} \right)^2 P^f(k) \left[1 + \frac{26}{21} (T_{\deltab}(k) -1)\deltab(z_f) + \left(\frac{G_2(k)}{2} - \frac{1501}{1323} - \left(\frac{26}{21}\right)^2 (T_{\delta_{\rm b}}(k)-1) \right)
\delta_{\rm b}(z_f)^2 \right],
     \label{eq:pk_est2}
\end{align}
\end{widetext}
where  
\begin{align}
    G_2(k) \equiv \frac{1}{P^f(k)} \left.\frac{\partial^2 P(k)}{\partial \deltab^2}\right|_{{\rm G},\deltab = 0}
\end{align}
and $1501/1323$ is the linear limit of $G_2(k)/2$ for the matter power spectrum~\citep{Wagner+15b}, derived in an Einstein-de Sitter cosmology.
We explicitly wrote down the expansion up to
the second order in $\deltab$ and show below that
the second order term is negligible for moderate values of $|\Omega_K|$ that are consistent with the current bounds, $|\Omega_K| \lesssim \mathcal{O}(10^{-1})$.
For the matter power spectrum, the normalized growth response $T_{\deltab}(k)$ ranges about $[0.3,1.6]$ at redshift $z \simeq [0,1.5]$ 
{in the range of $k \simeq [10^{-2}, 6] \, h {\rm Mpc}^{-1}$}
(see Fig.~2 of Ref.~\citep{Terasawa+22}) and $G_2(k)/2$ ranges about $[0.5,2]$ at $z=0$ 
{in the range of $k \simeq [10^{-2}, 2] \, h {\rm Mpc}^{-1}$}
(Fig.~2 of~\citep{Wagner+15a}). Since the prefactors of the first and second order terms before $\deltab$ and $\deltab^2$ are at most $\mathcal{O}(10^{-1})$, the contributions from the first and second order terms are at percent and sub-percent level,  respectively, even for a large curvature case of $|\Omega_K|\sim \mathcal{O}(10^{-1})$
(corresponding to $\deltab(t) \sim \mathcal{O}(10^{-1})$). As we will show below, the second or higher order terms are also negligible for the halo-matter or halo-auto power spectrum. Hence we consider the expansion up to the first order of $\delta_{\rm b}$ throughout this paper.

We also define the normalized growth response to $h$ within the flat model as:
\begin{align}
T_{h}^{\rm mm}(k)\equiv \left[2\frac{\partial \ln D}{\partial h}\right]^{-1}
\frac{\partial \ln P_{\rm mm}(k;h)}{\partial h}.
\label{eq:T^mm_h}
\end{align}
As shown numerically in Refs.~\citep{2014PhRvD..90j3530L, Terasawa+22} (also as will be shown in  Fig.~\ref{fig:Tmm-hm-hh}), these two responses agree well even in the nonlinear regime: $T_{\deltab}^{\rm mm}(k) \simeq T_h^{\rm mm}(k)$. Finally, using this approximate identity $T_{\deltab}(k) \simeq T_h(k)$ and ignoring the terms of second or higher order in $\deltab(t)$, we can approximate the estimator as
\begin{align}
    \tilde{P}(k,z;\Omega_K) &= \left(\frac{D(z)}{D_f(z_f)} \right)^2 P^f(k,z_f) \nonumber \\
    &\times \left[1 + \frac{26}{21} (T_{h}(k) -1)\deltab(z_f) \right].
     \label{eq:pk_estimator1}
\end{align}
By construction, the estimator reproduces the linear prediction at $k \rightarrow 0$, where the first order term vanishes because $T_h^{\rm mm}(k\rightarrow 0) = 1$.
The novel feature of the above estimator (Eq.~\ref{eq:pk_estimator1}) is that it allows one to compute the nonlinear matter power spectrum 
for nonflat universe from the quantities in flat universe. 
With this estimator we can extend emulators available in the community, which are applicable only for flat universes, to predict the power spectrum in a nonflat universe.

\subsection{SU approach for halo power spectra}
\begin{figure}
	\includegraphics[width=\columnwidth]{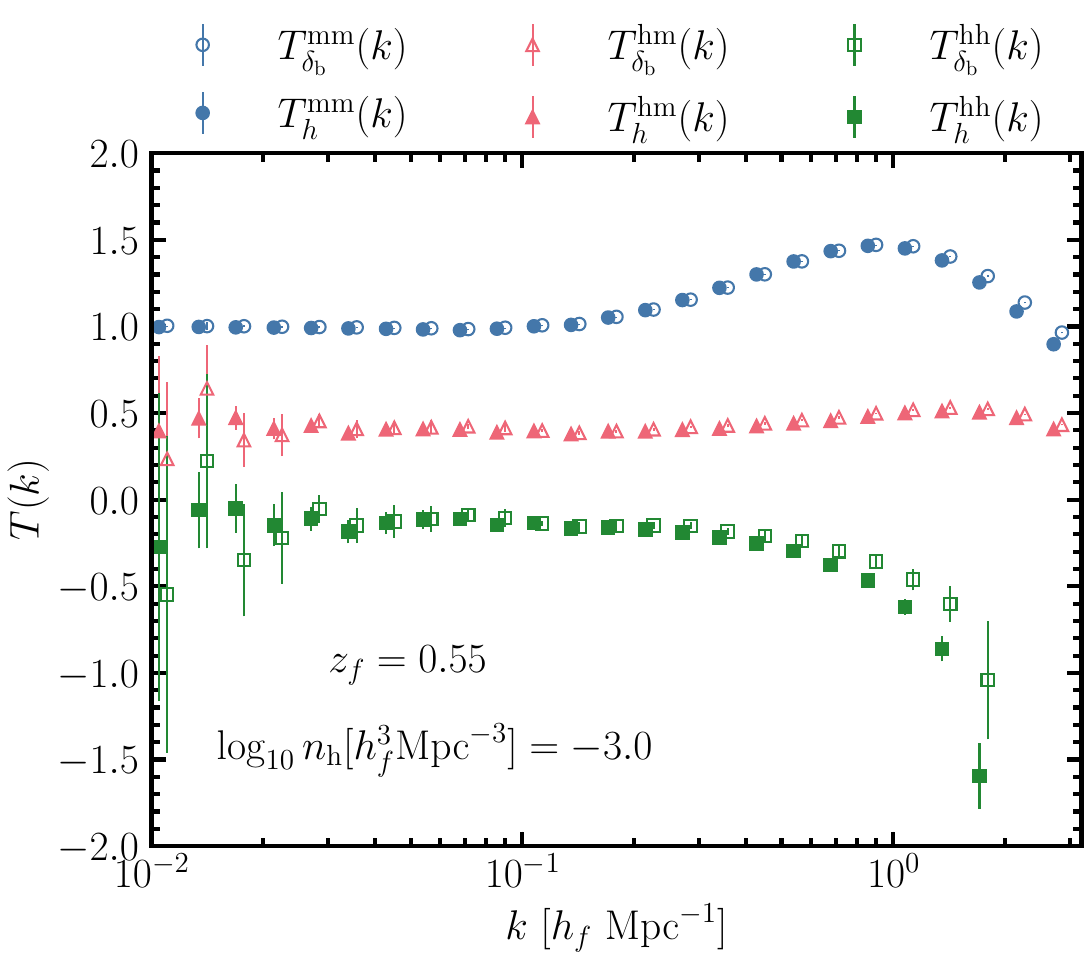}
\caption{The responses of the power spectra to $\deltab$ and $h$ measured from the $N$-body simulations (see Section \ref{sec:simulation} for details of the simulations). 
The circle, triangle, and square symbols show
the responses of the matter power spectrum ($T^{\rm mm}(k)$), halo-matter power spectrum ($T^{\rm hm}(k)$), and halo power spectrum ($T^{\rm hh}(k)$), respectively. 
For $T^{\rm hh}(k)$, we only plot the scales where halo power spectra after subtracting the shot noise have positive values. We slightly shift the symbols of $T_h(k)$ along the $x$-axis for illustration.
}
	\label{fig:Tmm-hm-hh}
\end{figure}
In this paper, we extend the SU approach mentioned above to the halo-matter and halo-auto power spectra.
Similarly to the above discussion,
we define $T_{\delta_{\rm b}}^{\rm XY}(k)$ and $T_{h}^{\rm XY}(k)$ as
\begin{align}
T_{\delta_{\rm b}}^{\rm XY}(k) &\equiv \left[2\frac{\partial \ln D(\deltab)}{\partial \delta_{\rm b}}\right]^{-1}
\left.\frac{\partial \ln P_{\rm XY}(k;\delta_{\rm b})}{\partial \delta_{\rm b}}\right|_{{\rm G}, \delta_{\rm b}=0}, \nonumber \\
T_{h}^{\rm XY}(k) &\equiv \left[2\frac{\partial \ln D}{\partial h}\right]^{-1}
\frac{\partial \ln P_{\rm XY}(k;h)}{\partial h},
\label{eq:normalized_response_db}
\end{align}
where $\rm{XY} = \{\rm mm,hm,hh \}$. When considering the response of $P_{\rm hm}(k)$ or $P_{\rm hh}(k)$, we perform the derivative keeping the comoving halo number density fixed.
As we will specify the latter in Sec.~\ref{appendix:pk_response}, these responses are not the same as the growth response in the SU approach where the derivative is performed keeping the halo mass threshold fixed.

As a highlight of our approach, in Fig.~\ref{fig:Tmm-hm-hh} we show the responses of power spectra of matter-matter, halo-matter, and halo-halo
to $\delta_{\rm b}$ or $h$, $T^{\rm mm}$, $T^{\rm hm}$ and $T^{\rm hh}$, which are computed using Eq.~(\ref{eq:normalized_response_db}). 
At the large-scale limit $k\rightarrow 0$, the matter-matter response $T^{\rm mm}_{\deltab}= 1$ and $T^{\rm mm}_h=1$ by definition. Given that the
abundance-matched halo samples correspond to the same initial density peaks, which is true as long as mergers do not severely affect this correspondence,  
the clustering amplitudes of the abundance-matched halos on large scales barely change. This is why $T^{\rm hh}(k \rightarrow 0) \simeq 0$ and 
$T^{\rm hm}$ is smaller than $T^{\rm mm}$.
The figure also shows that the responses for ``mm'', ``hm'' and ``hh''
have quite different 
$k$-dependence. Hence accurate calibration is important to capture the scale dependence.
The responses for $\deltab$ and $h$ are in good agreement with each other, for all the ``mm'', ``hm'' and ``hh'' power spectra, which validates our method to approximate the response to $\deltab$ by the response to $h$.

Finally, using the approximate identity $T_{\deltab}(k) \simeq T_h(k)$, we can obtain the estimator as
\begin{align}
    \tilde{P}_{\rm XY}(k,z;\Omega_K) &= \left(\frac{D(z)}{D_f(z_f)} \right)^2 P_{\rm XY}^f(k,z_f) \nonumber \\
    &\times \left[1 + \frac{26}{21} (T^{\rm XY}_{h}(k) -1)\deltab(z_f) \right].
     \label{eq:pk_estimator}
\end{align}

\section{Total response estimator using flat \LCDM~ model}\label{sec:SSC}

In this section, we discuss that our approximation using the response of the power spectrum to the Hubble parameter can be used to calibrate the SSC of a cosmological observable. Here the SSC is the sampling variance error of the observable in a finite-volume 
survey, which arises from the mode coupling of density fluctuations in the survey with density fluctuations on scales greater than the survey window, i.e. super-survey modes~\citep{2013PhRvD..87l3504T,2014PhRvD..89h3519L,2019MNRAS.482.4253T}. 
For instance, 
the SSC gives a dominant source of the sample variance on scales larger than the scales where the shot noise dominates, e.g., for the two-point correlation function of cosmic shear, which is given by the 
weighted line-of-sight integration of the matter power spectrum. 

SSC for the X- and Y-observables in the $i$-th and $j$-th bins, respectively, is generally expressed, as proposed by Ref.~\cite{2013PhRvD..87l3504T}, as
\begin{align}
    C^{({\rm XY}){\rm SSC}}_{ij} \approx \sigma_{\rm b}^2 \frac{\partial \mathcal{O}_{{\rm X}i}}{\partial \delta_{\rm b}} \frac{\partial \mathcal{O}_{{\rm Y}j}}{\partial \delta_{\rm b}},
\end{align}
where $\sigma_{\rm b}^2$ is the variance in the linear density filtered by
a window function assuming a sufficiently large survey volume in that the super-survey modes ($\delta_{\rm b}$)
are in the linear regime,
and $\partial {\cal O}/\partial \delta_{\rm b}$ is the {\it total} response of the observable to $\delta_{\rm b}$ including both the growth and dilation 
responses (see below).
The observables can be any statistical cosmological quantities such as power spectrum, bispectrum, and cluster mass function. 
In this paper, we consider, as $\mathcal{O}_{\rm X}$ and/or $\mathcal{O}_{\rm Y}$,  either of the matter, halo-matter, or halo-halo power spectum and the $i$-th index in the above equation corresponds to the $i$-th 
$k$ bin. Otherwise, the above equation is a general expression of SSC. We also note that, once the total response for the 3D observable (e.g. 
the matter power spectrum) is given, the SSC 
to the corresponding angular observable (e.g., cosmic power spectrum) can be obtained by a weighted line-of-sight integral of the 3D SSC term~\citep{2019MNRAS.482.4253T}.

In the following, we provide an approximate way to compute the total response using the power spectrum response to $h$.

\subsection{Preliminary}\label{sec:Preliminary}

First, we summarize analytical methods of computing the total response proposed in the literature.

\noindent$\bullet${\it Perturbation theory} --    
The total response can be computed using the perturbation theory (PT) \cite{2016JCAP...09..007B} as
\begin{align}
\frac{\partial \ln P_{\rm mm,PT}(k)}{\partial \delta_{\rm b}} &= \frac{68}{21}  - \frac{1}{3} \frac{\partial \ln k^3 P_L(k)}{\partial \ln k}, \nonumber \\
\frac{\partial \ln P_{\rm gm,PT}(k)}{\partial \delta_{\rm b}} &= \frac{68}{21} + \frac{b_{g,2}}{b_g} -b_g  - \frac{1}{3} \frac{\partial \ln k^3 P_L(k)}{\partial \ln k}, \nonumber \\
  \frac{\partial \ln P_{\rm gg,PT}(k)}{\partial \delta_{\rm b}} &= \frac{68}{21} +2\frac{b_{g,2}}{b_g} -2b_g  - \frac{1}{3} \frac{\partial \ln k^3 P_L(k)}{\partial \ln k},
  \label{eq:B16response}
\end{align}
where $b_g$ and $b_{g,2}$ denote the first- and second-order Eulerian galaxy biases.
Using the halo occupation distribution
(HOD) prescription~\citep[e.g.][]{2005ApJ...633..791Z}, $b_g$ is given as
\begin{align}
  b_g = \frac{1}{\bar{n}_g} \int~\mathrm{d}M \frac{\mathrm{d}n}{\mathrm{d}M}(M) b_{{\rm h},1}(M) \langle N_\mathrm{g}(M)\rangle,
\end{align}
with the mean number density of galaxies
\begin{align}
\bar{n}_g=\int~\mathrm{d}M \frac{\mathrm{d}n}{\mathrm{d}M}(M) \langle N_\mathrm{g}(M)\rangle,
\end{align}
where $\langle N_\mathrm{g}(M)\rangle$ is the mean number of galaxies in a halo,
$\mathrm{d}n/\mathrm{d}M$ is the halo mass function in the mass range 
$[M,M+\mathrm{d}M]$, and $b_{{\rm h},1}(M)$ is the linear bias of halos of mass $M$. 
Throughout the paper, we use the fitting function by Ref.~\cite{Tinker10} for the linear halo bias $b_{{\rm h},1}(M)$.

Similarly, $b_{g,2}$ is given as
\begin{align}
  b_{g,2} = \frac{1}{\bar{n}_g} \int\! \mathrm{d}M \frac{\mathrm{d}n}{\mathrm{d}M}(M) b_{{\rm h},2}(M) \langle N_\mathrm{g}(M)\rangle,
\end{align}
where $b_{{\rm h},2}(M)$ is the 2nd-order halo bias of halos of mass $M$. Throughout this paper, we use the fitting formula of $b_{{\rm h},2}$ proposed in Ref.~\cite{2017MNRAS.465.2225H}, which gives $b_{{\rm h},2}(M)$
in terms of the linear halo bias, i.e.
$b_{{\rm h},2} = b_{{\rm h},2}(b_{{\rm h},1})$.

As we will see in Figs.~\ref{fig:Pmm_total_resp} and \ref{fig:PgX_total_resp}, the perturbation theory well describes the response on large scale, but it fails to reproduce 
the small-scale behavior, 
where perturbation theory itself starts to break down.  

\noindent$\bullet${\it Halo Model} -- 
The halo model~\citep[e.g.][]{sheth02} is commonly used to calculate the response of the matter power spectrum in the literature~\citep{2013PhRvD..87l3504T,2014PhRvD..89h3519L,Chiang2014}. 
Ref.~\citep{cosmolike} derived the responses for the multiprobe power spectrum, denoted as 
$P_{\rm XY}$, based on the halo model. Following the notations in Ref.~\citep{cosmolike}, the 
responses are written as
\begin{align}
  \parfrac{P_{\rm XY}(k)}{\deltab} 
  &= \left(\frac{47}{21} + \frac{I_{\rm X}^2(k)}{I_{\rm X}^1(k)} + \frac{I_{\rm Y}^2(k)}{I_{\rm Y}^1(k)}  - 
  \frac{1}{3} \frac{\partial\ln P_L(k)}{\partial\ln k}\right) \nonumber \\
  &\times I_{\rm X}^1(k)I_{\rm Y}^1(k) P_L(k)  \nonumber \\
  &+ I_{\rm XY}^1(k,k) - \left[b_{\rm X,X=g} + b_{\rm Y,Y=g} \right]P_{\rm XY}(k),
  \label{eq:HMresponse}
\end{align}
where $b_{{\rm X,X}={\rm g}} = b_{\rm g}$ for ${\rm X}={\rm g}$ and otherwise $b_{{\rm X,X}={\rm g}} = 0$ 
and the functions $I_{\rm X}^\alpha$ and $I_{\rm XY}^\alpha$ are defined as
\begin{align}
    I_{\rm X}^{\alpha}(k) &= \int\!\mathrm{d}M \frac{\mathrm{d}n}{\mathrm{d}M}(M) b_{\rm{h},\alpha}(M) \tilde{u}_{\rm X}(k;M), \nonumber \\ 
    I_{\rm XY}^{\alpha}(k,k') &= \int\!\mathrm{d}M \frac{\mathrm{d}n}{\mathrm{d}M}(M) b_{\rm{h},\alpha}(M) \tilde{u}_{\rm X}(k;M)\tilde{u}_{\rm Y}(k';M),
    \label{eq:HM_I_X_definition}
\end{align}
where  $\tilde{u}_{\rm X}(k;M)$ is the Fourier transform of the radial profile of tracers X in host halos, multiplied by the number density normalization and the HOD function for galaxies following the notations in Ref.~\cite{cosmolike}. Note that  $\tilde{u}_{\rm X}(k;M)$ has a dimension of  volume.

We stress that we included the terms proportional to $I_{\rm X}^2(k)$ in the parentheses of Eq.~(\ref{eq:HMresponse}), which were missing 
in Ref.~\citep{cosmolike}. These terms depend on the 2nd-order halo bias, $b_{{\rm h},2}$, which arises from the response of the linear halo bias 
$b_{{\rm h},1}$ to $\deltab$. For the matter power spectrum (for the case ${\rm X} = {\rm Y} ={\rm m}$),
the terms proportional to $I_{\rm m}^2(k)$ are negligible due to the halo model consistency relation \citep{Wagner+15b}.
Since $I_g^1(k) \rightarrow b_g$ and  $I_g^2(k) \rightarrow b_{g,2}$ at the $k \rightarrow 0$ limit, $I_g^2(k)/I_g^1(k) \rightarrow b_{g,2}/b_g$ at
the limit, which reproduces the PT response (Eq.~\ref{eq:B16response}).
Thus we believe that the above response formula (Eq.~\ref{eq:HMresponse}) is more accurate in the sense that it includes the PT theory at the limit 
of $k\rightarrow 0$.

We will below assess the accuracy and limitation of these analytical formulae of the power spectrum responses by comparing the model predictions with 
the simulation results.

\subsection{$h$ response method}
The long-wavelength modes (super-survey modes) whose
wavelengths are larger than the survey volume/simulation box affect the growth of LSS via mode coupling. The effects of the long-wavelength modes can be considered as the background density (mean density) modulation, $\deltab$, which in turn can be interpreted as the local effective cosmology with non-zero curvature corresponding to $\deltab$ (the SU approach~\citep[e.g.][]{2011JCAP...10..031B,2013PhRvD..87l3504T,2014PhRvD..89h3519L,2014PhRvD..90j3530L,Wagner+15a}).

Different large-scale structure tracers are measured with respect to either the {\it global} or {\it local} mean density of the tracers, where 
the {\it local} mean is the average density of the tracers in a finite-volume survey. Depending on this difference,
the response of power spectrum of X and Y tracers,
$P_{\rm XY}$, to the super survey modes can be decomposed into three contributions~\citep[e.g.][]{2014PhRvD..89h3519L, 2019MNRAS.482.4253T}:
\begin{align} 
  \left.\frac{\partial \ln P_{\rm XY}(k;\delta_{\rm b})}{\partial \delta_{\rm b}}\right|_{\rm total} &= n + \left.\frac{\partial \ln P_{\rm XY}(k;\delta_{\rm b})}{\partial \delta_{\rm b}}\right|_{\rm G} \nonumber \\
  &-\frac{1}{3} \frac{\partial \ln k^3 P_{\rm XY}(k)}{\partial \ln k}. 
  \label{eq:p_XY_total}
\end{align}
The first term on the r.h.s. accounts for the change in the mean density of the tracers by $\deltab$, used in the definition of the density fluctuation field; 
for example, for the matter field (XY=mm), $\delta_\mathrm{m}=\rho_\mathrm{m}/\bar{\rho}_\mathrm{m}-1$. 
The cases of $n = 2,1$ and $0$ correspond to 
``${\rm XY}$''$=$``${\rm mm}$'', ``${\rm gm}$'' and ``{\rm gg}'', respectively, which are 
relevant to cosmic shear, galaxy-galaxy weak lensing and 
galaxy-galaxy clustering, respectively \citep{2019MNRAS.482.4253T}.
The second term is the growth response, which describes the fractional change in the power spectrum amplitude by
the presence of $\deltab$. 
The last term is the dilation response, which originates from the change in the physical scale corresponding to a given comoving scale due to the change in cosmic expansion.

Using this decomposition, we propose the matter power spectrum response estimator as
\begin{align} 
  \left.\frac{\partial \ln P_{\rm mm}(k;\delta_{\rm b})}{\partial \delta_{\rm b}}\right|_{\rm total} &= 2 + \frac{26}{21} T_h^{\rm mm}(k)
  -\frac{1}{3} \frac{\partial \ln k^3 P_{\rm mm}(k)}{\partial \ln k}, \nonumber \\
  \label{eq:p_mm_total}
\end{align}
where we have used the fact that the normalized growth response to $\deltab$ is approximated by that to $h$:
\begin{align}
\left.\frac{\partial \ln P_{\rm mm}(k;\delta_{\rm b})}{\partial \delta_{\rm b}}\right|_{G} = \frac{26}{21} T_{\deltab}^{\rm mm}(k) \simeq
\frac{26}{21} T_{h}^{\rm mm}(k).
\label{eq:growth_response_approx}
\end{align}
Similarly, as we will show in Appendix~\ref{appendix:pk_response}, we can use the approximate identity of the responses of halo-matter and halo-auto power spectra to predict the growth responses of galaxy-matter and galaxy-auto power spectra with the HOD description:
\begin{widetext}
\begin{align}
  \left.\parfrac{P_{\rm gm}(k)}{\deltab}\right|_{\rm G}
 &\simeq \frac{1}{\bar{n}_g}\int \mathrm{d}M \left\{ \dfrac{\mathrm{d}\langle \tilde{U}_g (k;M)\rangle}{\mathrm{d}M} n(>M) \right.
  \left[ b_{\mathrm{h}, 1}^L(>M)P_{\rm hm}(k;M) 
  +\frac{26}{21} T_{h}^{\rm hm}(k;>M) P_{\rm hm}(k;>M) \right]  \nonumber \\
  &~~~~~\left. + \frac{1}{3} P_\mathrm{hm}(k;>M)
\frac{\partial}{\partial \ln k} \frac{\mathrm{d}\ln \langle \tilde{U}_g(k;M)\rangle}{\mathrm{d}M} \right\} - b_g^L P_{\rm gm}(k), 
  \label{eq:G^gm} 
\end{align}
\begin{align}
  \left.\parfrac{P_{\rm gg}(k)}{\deltab}\right|_{\rm G} 
  &\simeq \frac{1}{\bar{n}_g^2} \iint \mathrm{d}M \mathrm{d}M' \left\{ \frac{\mathrm{d}\langle \tilde{U}_g (k;M)\rangle}{\mathrm{d}M} \frac{\mathrm{d}\tilde{U}_g(k;M')}{\mathrm{d}M'} \right. 
  n(>M) n(>M') 
  \left[ 2b_{\mathrm{h}, 1}^L(>M') P_{\rm hh}(k;>M,M') \right. \nonumber \\
  &~~~~~ + \left.\left. \left(\frac{26}{21}\right) T_{h}^{\rm hh}(k;>M,>M') 
   P_{\rm hh}(k;>M,>M') + \frac{2}{3} {P_{\rm hh}(k;>M,>M')}
\frac{\partial}{\partial \ln k} \frac{\mathrm{d} \ln \langle \tilde{U}_g(k;M)\rangle}{\mathrm{d}M}
     \right] \right\}  \nonumber \\
  &~~~~~ + \frac{1}{\bar{n}_g^2} \int \mathrm{d}M
  \frac{\mathrm{d}n}{\mathrm{d}M}(M)  \left[ b_{\mathrm{h}, 1}^L(M) \langle\tilde{U}_g^2(k;M)\rangle + \frac{1}{3}\frac{\partial \langle\tilde{U}_g^2(k;M)\rangle}{\partial \ln k} 
  \right] - 2b_g^L P_{\rm gg}(k),
  \label{eq:G^gg}
\end{align}
\end{widetext}
where the super-script $L$ denotes the Lagrangian bias and $\tilde{U}_g(k;M)$ describes the quantity arising from 
the galaxy-halo connection (see Appendix~\ref{sec:derivation} for the definition).

Since the $h$ response is defined within flat cosmologies, we can predict the power spectrum response from flat cosmology predictions or simulations without 
performing a pair of nonflat universe simulations {following the SU framework}.
As we will show below, using the simulation-based emulator we can predict the response accurately down to the nonlinear scale.

Finally, to ensure the estimator is correct at large scales, we stitch the estimator with perturbation theory prediction~\citep{2016JCAP...09..007B} described above in Eq.~(\ref{eq:B16response}).
Specifically, we smoothly stitch the estimator with perturbation theory prediction as
\begin{align}
  \frac{\partial \ln P_{\rm XY}(k)}{\partial \delta_{\rm b}} &= \frac{\partial \ln P_{\rm XY,PT}(k)}{\partial \delta_{\rm b}}e^{-(k/k_{\rm switch})} \nonumber \\
  &+ \frac{\partial \ln P_{\rm XY,est}(k)}{\partial \delta_{\rm b}} \left[1 - e^{-(k/k_{\rm switch)}} \right].
  \label{eq:stitch}
\end{align}
Throughout the paper, we adopt the switching scale $k_{\rm switch} = 0.08~h~{\rm Mpc}^{-1}$ for XY = mm, gm and gg.

\section{Simulations}\label{sec:simulation}

	\begin{table*}
		\caption{Details of $N$-body simulations for different cosmological models. 
        The columns ``$\Omega_K$'' and ``$h$'' give their values of the curvature parameter and Hubble parameter
        that are  
        employed in each simulation, while we fix 
        other cosmological parameters $\{\omega_{\rm c},\omega_{\rm b}, A_s, n_s\}$, which are needed to specify the linear power spectrum for the initial conditions, to the values for the fiducial 
        {\it Planck} cosmology (see text for details). 
        $\Omega_{\rm m}$ and $\Omega_\Lambda$ are specified by a given set of $\Omega_K$ and $h$, 
        because we keep 
        $\Omega_{\rm m}h^2$ fixed and $\Omega_{\Lambda}=1-\Omega_{\rm m}-\Omega_{\rm K}$.
        The column ``$N_{\rm real}$'' denotes the number of realizations, with different initial seeds, used for each model. 
        We employ the ``paired-and-fixed'' method in~\cite{2016MNRAS.462L...1A} to reduce the sample variance effect in small $k$ bins for the power spectrum measurement: it uses the paired (2) simulations by design (see text for details). 
        The column ``redshift ($z$)''
        gives the redshifts of simulation outputs: for $\KCDM$ model, we properly choose the redshifts corresponding to the same cosmic time for each of redshifts, $z=\{0.0,0.549,1.025,1.476\}$ in  the ``fiducial'' model in the SU approach (see around Eq.~\ref{eq:scale_factor_mapping} in Section~\ref{sec:SUapproach}).  
        All the simulations are done in the fixed comoving box size without $h$ in its units, i.e.\ $L\simeq 2.97\,{\rm Gpc}$ (corresponding to $2\,h_f^{-1}{\rm Gpc}$ for the fiducial model)
        and with the same particle number, i.e.\ $N_p=2048^3$.}
        \label{tab:simulations}
		\begin{center}
		\begin{tabular}{l|cccl}\hline\hline 
		Name & $\Omega_K$ & $h$ &  $N_{\rm real}$    & redshift ($z$) \\ \hline 
		flat (fiducial) & 0 & $0.6727$ &  2  & $\{0.0, 0.549, 1.025, 1.476\}$ \\ \hline
		$\Omega_K$-$\Lambda$CDM1 & 0.00663 & 0.6749 & 8  &$\{-0.0033,  0.544,  1.018,  1.467\}$ \\
		& $-0.00672$ & 0.6705  & 8   &$\{0.0033, 0.554, 1.031, 1.484\}$ \\ \hline 
        $\Omega_K$-$\Lambda$CDM2 & 0.1 & 0.7091  &2  & $\{-0.059,    0.482,  0.955,  1.405\}$ \\
        & $-0.1$ &0.6414  &2  &  $\{0.043, 0.600,  1.079,  1.531\}$\\ 
        \hline 
		{$h$}-$\Lambda$CDM & 0 & 0.6927 & 8   &$\{0.0, 0.549, 1.025, 1.476\}$\\ 
		& 0& 0.6527 & 8   &$\{0.0, 0.549, 1.025, 1.476\}$ \\ \hline\hline
		\end{tabular}\\
		\end{center}
		\end{table*}

\subsection{$N$-body simulations}

In this section we give a brief summary of the simulations used in this paper. 
Our simulations follow the method in~\citet{Nishimichi_2019}. 

We use {\tt Gadget-2}~\citep{gadget2} to carry out $N$-body simulation for a given cosmological model. 
The initial conditions are set up at redshift $z_i = 29$
using the second-order Lagrangian perturbation theory~\citep{scoccimarro98,crocce06b} implemented by~\citet{nishimichi09}
and then parallelized in \citet{Valageas11a}. 
We use the public code {\tt CAMB}~\citep{camb} to compute 
the transfer function for a given model, which is used 
to compute the input linear power spectrum.
For all simulations in this paper, 
we use the same simulation box size in Gpc (i.e.\ without $h$ in the units) and the same number of particles:  
$L = 2\,h_f^{-1}\mathrm{Gpc}\simeq 2.97\,{\rm Gpc}$ (without $h$ in units)
and $N_p=2048^3$, which correspond to the particle Nyquist wavenumber, $k =  3.2~ h_f \rm{Mpc}^{-1}$.
In the following we will show the results at wavenumbers smaller than this Nyquist wavenumber. 
In this paper we use simulations for 4 different cosmological models, denoted as ``fiducial'' flat $\Lambda$CDM, 
``$\KCDM$1'', ``$\KCDM$2'', and ``$h$-$\Lambda$CDM'' models, respectively, as given 
in Table~\ref{tab:simulations}. Here the cosmological parameters for the ``fiducial'' model are chosen to be consistent with those
for the {\it Planck} 2015 best-fit cosmology~\cite{planck-collaboration:2015fj}. The cosmological parameters for each of the nonflat cosmological models 
are chosen so that it has the fiducial $\Lambda$CDM model as the corresponding flat $\Lambda$CDM model in the SU approach. 
We use paired simulations for ``$\KCDM$1'' model to compute the power spectrum response 
with respect to $\delta_{\rm b}$ ($T_{\delta_{\rm b}}$), where the curvature parameters are specified by 
$\delta_{\rm b}=\pm 0.01$ at $z_f=0$. The ``$h$-$\Lambda$CDM'' model is for computing the response with respect to $h$ ($T_h$): 
here, we chose a step size of $\delta h= \pm 0.02$ for the numerical derivative.
We also use the simulations for nonflat $\Lambda$CDM models with 
$\Omega_K= \pm 0.1$, named as ``$\KCDM$2'',  
to assess how our method can approximate the halo-matter and halo-auto power spectrum for nonflat models.

Table~\ref{tab:simulations} gives the values of $\Omega_K$ and $h$, and we use the fixed values of other cosmological parameters, given as $(\omega_{\rm c},\omega_{\rm b},A_s, n_s)=(0.1198,0.02225,2.2065 \times 10^{-9}, 0.9645)$, which 
specify the transfer function and the primordial power spectrum, or equivalently the linear matter power spectrum. Note that we 
also include the effect of massive neutrinos on the linear matter power spectrum, assuming 
$\Omega_\nu h^2=0.00064$ corresponding to $m_{\nu,{\rm tot}}=0.06\,{\rm eV}$, the lower limit inferred from the
oscillation experiments~\citep[see Ref.][for details]{Nishimichi_2019}. Hence the physical density parameter of total matter is 
$\Omega_{\rm m}h^2=\omega_{\rm c}+\omega_{\rm b}+\omega_\nu$.
Note that $\Omega_{\rm m}$ and
$\Omega_\Lambda$ are specified 
by a given set of the parameters for each model: $\Omega_{\rm m}=\Omega_{{\rm m}f}h_f^2/h^2$ and 
$\Omega_\Lambda=1-\Omega_{\rm m}-\Omega_{K})$. 
For each model, we use the outputs at 4 redshifts, $z_f\simeq 0, 0.55, 1.03$ and $1.48$. Since the ``fiducial''
flat $\Lambda$CDM model is the flat model in the SU method, each redshift 
for the fiducial flat model corresponds to a slightly different redshift in each nonflat model, which 
is computed from Eq.~(\ref{eq:scale_factor_mapping}).

Furthermore, we use simulations that are run using the ``paired-and-fixed'' method in~\citet{2016MNRAS.462L...1A}, where 
the initial density field in each Fourier mode is generated from the fixed amplitude of the power spectrum 
$\sqrt{P(k)}$ and the paired simulations with reverse phases, i.e.\ $\delta_{\bk}$ and $-\delta_{\bk}$. The mean power spectrum of the paired runs fairly well reproduces 
the ensemble average of many realizations even in the nonlinear regime~\cite{2016MNRAS.462L...1A,2018ApJ...867..137V}. The paired-and-fixed simulations allow us to significantly reduce the sample variance in the power spectrum estimation. 
``2 (8)'' on the column $N_{\rm real}$ denotes one (four) pair(s) of the paired-and-fixed simulations. 
For all simulations, halos are identified using {\tt Rockstar}~\citep{Behroozi2013_}.
We adopt $M \equiv M_{\Delta} = 4\pi/3(R_{\Delta})^3 (\Delta\bar{\rho}_{\rm m0})$ with $\Delta = 200$ throughout the paper.
Note that when measuring the total response in the \KCDM1~simulations, we use the spherical overdensity $\Delta_W = 200/(1+\delta_{\rm b}(t))$ so that halos are identified using the same {physical} overdensity as in the corresponding global universe~\citep[e.g.][]{2016PhRvD..93f3507L}.
Note that all the $N$-body simulations for different cosmological models have a fixed mass resolution because they share the same $\Omega_{\rm m} h^2$ and comoving volume.

After we identified halo candidates, we determine whether they are 
central or satellite halos. When the separation of two different halos (between their centers) is closer than $R_{\rm 200m}$ of the more massive one, we mark 
the less massive one as a satellite halo. In the following, we use only central halos.

\subsection{Mock catalogs of galaxies in \KCDM~simulations}
\begin{table}
  \begin{center}
    \caption{The parameters of halo occupation distribution (HOD, Eq.~\ref{eq:HOD}) used for making the galaxy mock catalogs in this paper.}
    \vspace{0.3cm}
    \begin{tabular}{|c|c|} \hline 
      $\log M_{\rm min}[M_{\odot}/h_f]$    & $13.94$     \\
      $\sigma_{\ln{M}}$  & $0.5$  \\
      $\log M_0[M_{\odot}/h_f]$    & $13.72$   \\
      $\log M_1[M_{\odot}/h_f]$            & $14.46$   \\
      $\alpha_{\rm sat}$   & $1.192$  \\ 
      \hline
    \end{tabular}
    \label{tab:HOD_parameters}
  \end{center}
\end{table}

To validate our method to compute the galaxy clustering observables such as
galaxy-matter and galaxy-auto power spectra in nonflat cosmology and their total response, we build the galaxy mock catalogs using the halo catalogs of \KCDM~ simulations. 
We use the halo catalogs of \KCDM1 simulations
with the halo mass defined using $\Delta_W = 200/(1 + \deltab(t))$ to measure the total response, whereas we use the halo catalogs of \KCDM2 simulations
with the halo mass defined using $\Delta = 200$ to measure the galaxy clustering observables in the nonflat universe.
We assume that the galaxy profile around halos and the mean HOD depend only on halo mass and
are invariant to
cosmology, and populate galaxies into halos of each realization assuming the same HOD 
{in each \KCDM~ simulations.}
Specifically, we adopt the following central and satellite HODs \citep{2005ApJ...633..791Z}:
\begin{align}
  \langle N_c(M) \rangle &= \frac{1}{2} \left[1 + {\rm erf} \left(\frac{\log M - \log M_{\rm min}}{\sigma_{\ln M}} \right) \right], \nonumber \\
  \langle N_S(M) \rangle &= \Theta(M-M_0) \left(\frac{M-M_0}{M_1}\right) ^{\alpha_{\rm sat}},
  \label{eq:HOD}
\end{align}
where $\Theta$ is the step function. The total galaxy occupation is written as
\begin{align}
  \langle N_g(M) \rangle = \langle N_c(M) \rangle \left[1 + \langle N_s(M) \rangle \right].
\end{align}
We populate the central galaxies into the center of halos according to the Bernoulli distribution with mean $\langle N_c(M) \rangle $. For satellite galaxies, we populate them only into host halos each of which already hosts a central galaxy,
assuming they obey the Poisson distribution with mean $\langle N_S(M) \rangle $. 
We assume that the satellite galaxy density profile $u_S(r,M)$ follows the NFW profile~\citep{NFW} with mass-concentration relation $c(M)$ in Ref.~\cite{2015ApJ...799..108D}. We employ
the HOD parameters as denoted in Table~\ref{tab:HOD_parameters}.
These values correspond to the fiducial vaules of SDSS ``CMASS1'' sample in Ref.~\cite{HSCY1_Miyatake_validation} except for $\sigma_{\log{M}}$, for which we use $\sigma_{\log{M}} = 0.5$ instead of the fiducial value $\sigma_{\log{M}} = 0.7919$.

As noted in the previous section, each pair of the SU simulation uses the same initial seeds to reduce the sample variance. However, as we populated galaxies randomly, the reduction of the sample variance is partially ruined. Hence, we marginalized over 10 HOD seeds for each realization of $N$-body simulation to obtain the converged results.

\begin{figure*}
	\includegraphics[width=\columnwidth]{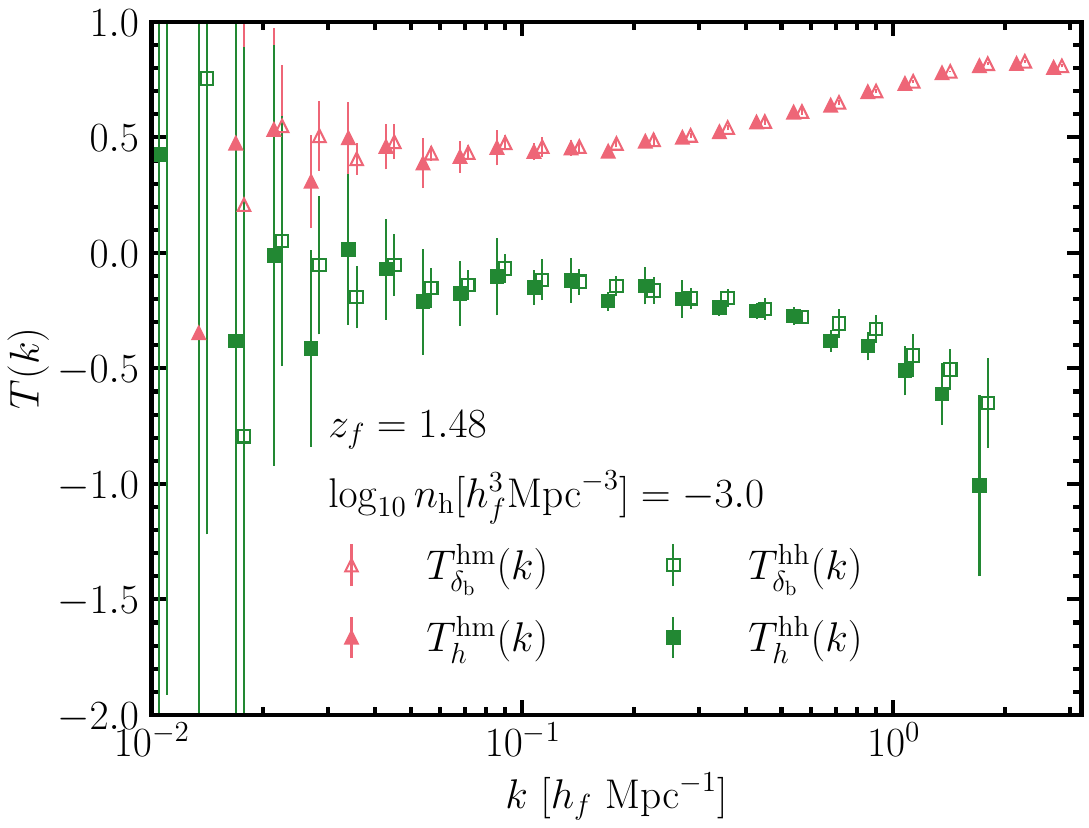}
    \includegraphics[width=\columnwidth]{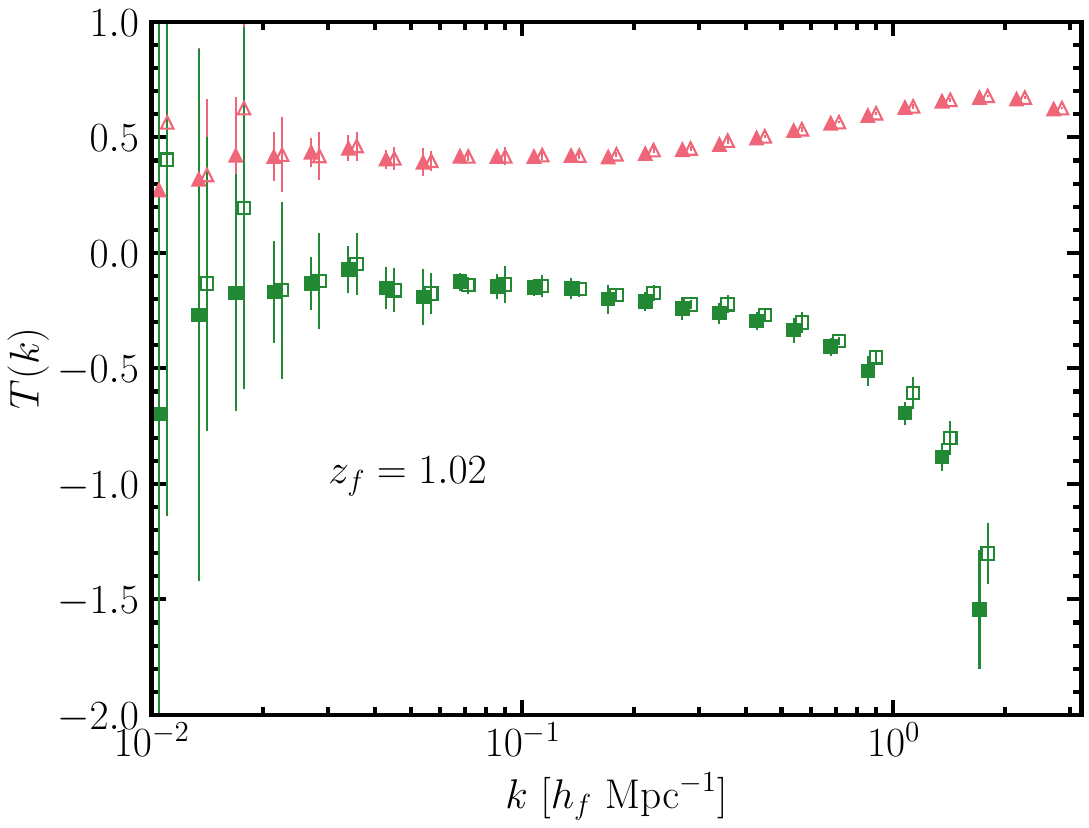}
    \includegraphics[width=\columnwidth]{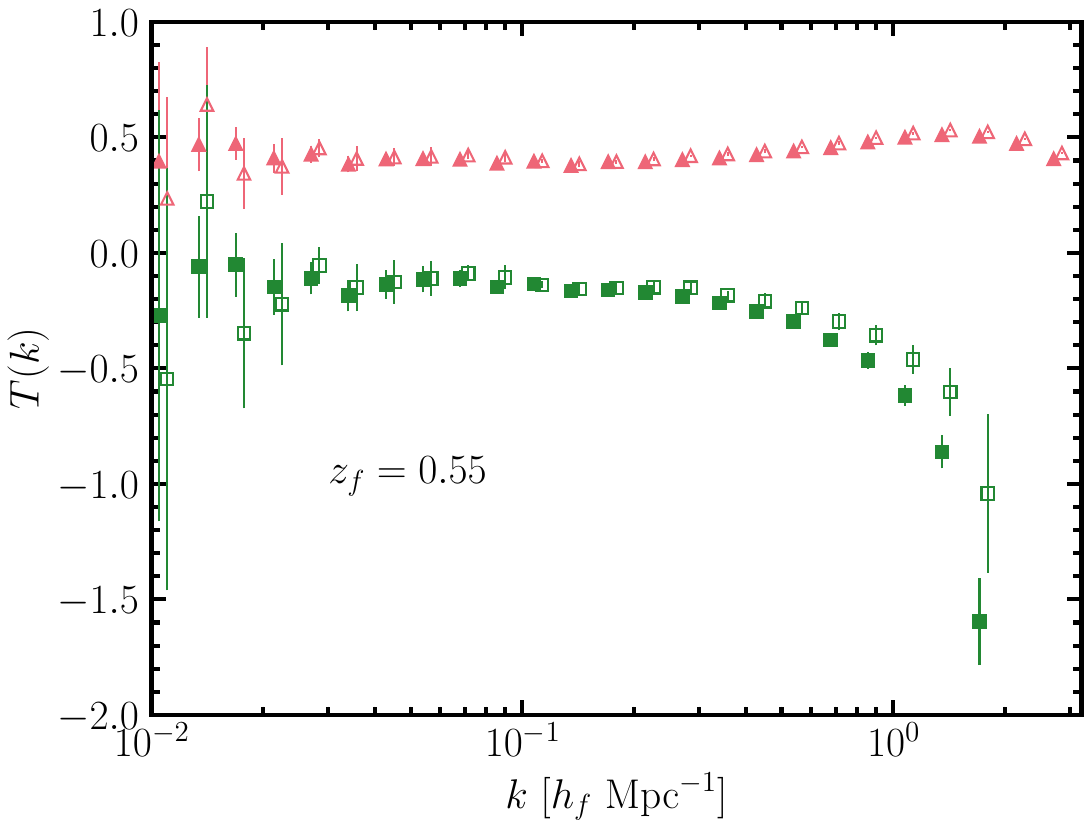}
    \includegraphics[width=\columnwidth]{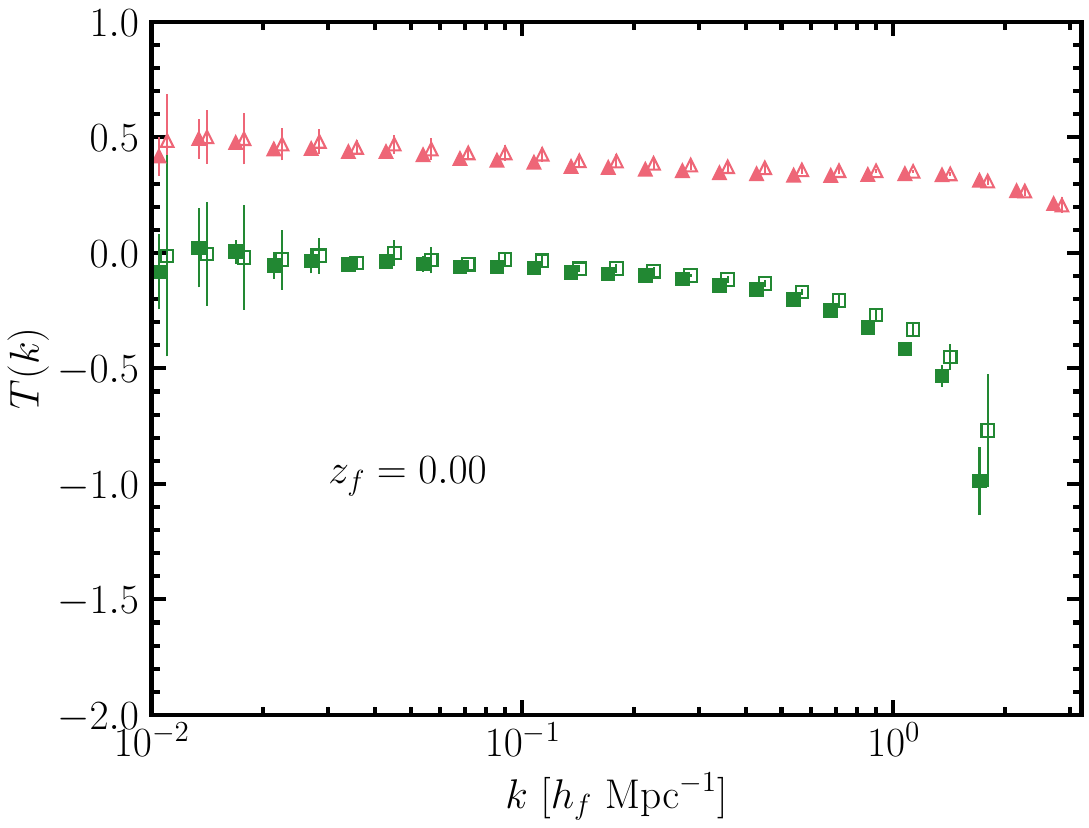}

\caption{Growth responses of halo-matter and halo-auto power spectra with respect to $\deltab$ ($h$), $T_{\deltab}^{\rm hm}(k)$ ($T_h^{\rm hm}(k)$) and $T_{\deltab}^{\rm hh}(k)$ ($T_h^{\rm hh}(k)$), at the four redshifts as denoted by the legend in each panel. We use 8 paired simulations for $\KCDM1$ and $h$-$\Lambda$CDM models in Table~\ref{tab:simulations} to compute these responses. 
    The symbols denote the mean of $T_{\delta_{\rm b}}$ or $T_h$ 
    in each $k$ bin, and the error bars (although not visible in some $k$ bins) denote the statistical errors for the simulation box with side length $L=2\,h_f^{-1}{\rm Gpc}$, which 
    are estimated from the standard deviations among the 8 paired simulations. Note that the range of $y$-axis is different in different panels.
    For $T^{\rm hh}(k)$, we only plot the scales where halo power spectra after subtracting the shot noise have positive values. We slightly shift the symbols of $T_h(k)$ along the $x$-axis for illustration.
}
	\label{fig:halo-matter-Tbh-nth}
\end{figure*}

\subsection{Measurements of power spectra and the responses}
\label{sec:sim_powerspectrum}

To calculate the power spectrum from each simulation output, we assign the $N$-body particles, halos, or galaxies on $2048^3$ grids using 
the cloud-in-cells (CIC) method~\citep{hockney81} to obtain the density fields of matter, halo or galaxy. After performing the Fourier transform, 
we correct for the window function of CIC following the method described in {Ref.~\cite{jing05}}.
We will show the results at wavenumbers smaller than the Nyquist frequency, $k = 3.2~h_f{\mathrm{Mpc}^{-1}}$. 
Furthermore, to evaluate the power spectrum at $k \ge 1.6~h_f{\mathrm{Mpc}^{-1}}$ accurately, we fold the particle positions
into a smaller box by replacing $\bold{x} \rightarrow \bold{x}\%(L/2)$, where the operation $a\% b$ stands for
the remainder of the division of $a$ by $b$. This procedure leads to effectively $2$ times higher resolution. 
For the halo-auto and galaxy-auto power spectrum, we subtract the shot noise from the measurements, 
where we simply assume the shot noise to be given by the number density of the tracers.

Since we use the fixed box size and the same particle number, we 
use the same $k$ binning to estimate the average of $|\delta_{\bf k}|^2$ in each $k$ bin to estimate the band power. 
We then use the two-side numerical derivative method to compute the power spectrum responses. 
We compute the growth response as
\begin{align}
	&\left.\frac{\partial \ln P(k,z)}{\partial \deltab}\right|_{{\rm G}, \rm sim} \nonumber\\
	&= \frac{\ln P_{+}(k;\delta_{\rm{b}0} = +\epsilon) - \ln P_{-}(k;\delta_{\rm{b}0} = -\epsilon) }{2\epsilon \times D(z)} 
 \label{eq:groath_resp_sim}
\end{align}
and compute the total response as
\begin{align}
	&\left.\frac{\partial P(k,z)}{\partial \deltab}\right|_{\rm total,sim} = (n-1)P(k,\delta_{\rm b0}=0) \nonumber \\
 &+ \frac{ P_{+}((1-\deltab/3)k;\delta_{\rm{b}0} = +\epsilon) -  P_{-}((1 + \deltab/3)k;\delta_{\rm{b}0} = -\epsilon) }{2\epsilon \times D(z)},
 \label{eq:total_resp_sim}
\end{align}
where $\delta_{\rm{b}0} \equiv \deltab(z_f=0)$ and $\epsilon=0.01$ for the \KCDM1 simulations.
To reduce statistical stochasticity (or sample variance), we employ the same initial seeds as those for the ``fiducial'' model. The column ``$N_{\rm real}$'' 
in Table~\ref{tab:simulations} denotes the number of realizations for paired simulations, where each pair uses the same initial seeds. 
For $\KCDM$1 and $h$-$\Lambda$CDM models, we run 8 paired simulations to estimate the statistical scatters.

\section{Results}\label{sec:results}

\subsection{Growth response}
\label{subsec:Growth response}
\begin{figure}
	\includegraphics[width=\columnwidth]{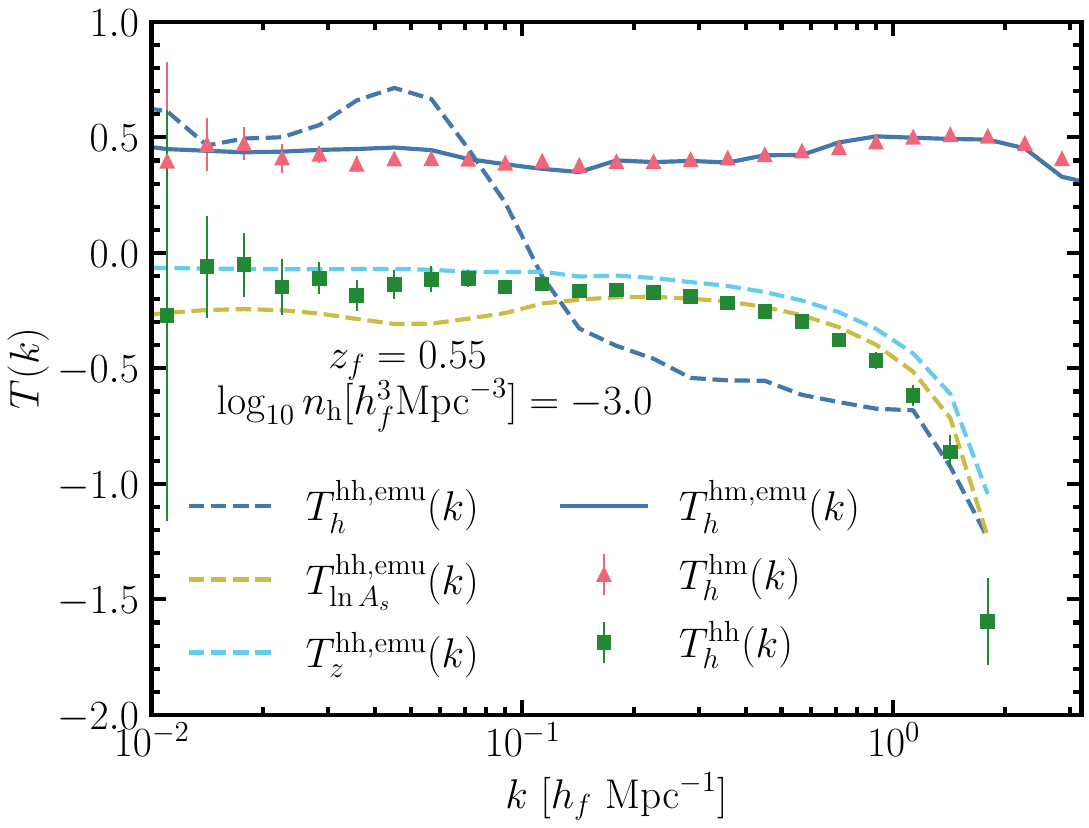}
\caption{Comparison of the simulation result for $T_{h}^{\rm hm}(k)$, $T_{h}^{\rm hh}(k)$ at $z_f=0.55$ with those computed using {\tt Dark Emulator}~\citep{Nishimichi_2019}, $T_{h}^{\rm hm, emu}(k)$, $T_{h}^{\rm hh, emu}(k)$.
}
	\label{fig:halo-matteer-DE-Tbh-nth}
\end{figure}

First, we study
the approximate identity of the growth response $T_{\deltab}(k) \approx T_h(k)$ with our $N$-body simulations.

In Fig.~\ref{fig:halo-matter-Tbh-nth}, we compare the growth response of halo-matter power spectrum to $\deltab$ and $h$, for the abundance-matched halo samples, at the four redshifts as in Table~\ref{tab:simulations}. These responses are calculated from $N$-body simulations for \KCDM1 and $h$-$\Lambda$CDM models. We can see that the approximate identity $T_{\deltab}^{\rm hm}(k) \approx T_{h}(k)$ holds for all four redshifts and down to the nonlinear scale.
Note that halos are selected using the abundance matching method, where the same number of halos, selected in the ascending order of masses, are identified 
in the two simulations when computing the responses.

In Fig.~\ref{fig:halo-matter-Tbh-nth}, we also compare the growth response of the halo-auto power spectrum.
Again it is clear that the approximate identities $T_{\deltab}^{\rm hh}(k) \approx T_{h} ^{\rm hh}(k)$ hold for all four redshifts and down to mildly nonlinear scales. 
Since the pairs of SU simulations (``\KCDM1") share the same Gaussian initial condition, the 
abundance-matched halo samples correspond to the same initial density peaks unless mergers severely affect this correspondence. Hence the pairs of SU simulations have the similar clustering amplitudes of the abundance-matched halos on large scales, which lead to $T^{\rm hh}(k \rightarrow 0) \simeq 0$. 
The large change in the response $T^{\rm hh}(k)$ around $k \sim 1~h_f~\rm{Mpc}^{-1}$ can be attributed to the exclusion effects of the halos, whereas the scales smaller than the exclusion scales does not contribute to the galaxy-auto power spectrum.

We showed the results for only one case of the number density selected halo sample,
but we confirmed the approximate identity also holds for other number densities.

As discussed in Ref.~\cite{Terasawa+22}, we expect the responses $T_{\deltab}(k)$ and $T_h(k)$ agree when the power spectrum is a functional of the amplitude of the linear power spectrum.
If we assume the universal halo mass function, which depends on cosmology only through $P_L(k)$, the corresponding halo bias is also determined only by $P_L(k)$. 
Hence, we can expect the responses $T(k)$ to the different parameters that leave the shape of the linear power spectrum unchanged agree with each other for the linear or quasi-nonlinear scales, where bias expansion of the density field is valid. Especially, on large scales, where 
$P_{\rm hm}(k) = b_1 P_L(k)$ ($P_{\rm hh}(k) = b_1^2 P_L(k)$) holds, the response $T(k)$ is 
expected to be constant.
On the other hand, the responses in the nonlinear regime could be different due to the difference in the growth history of the structure~\citep{Terasawa+22}, which leads to a change in the concentration of halos~\citep{2014PhRvD..90j3530L}. 
For $T_{\deltab}(k)$ and $T_h(k)$, we found out they agree with each other even in the nonlinear scale by numerical simulations.

In Fig.~\ref{fig:halo-matteer-DE-Tbh-nth} we assess the accuracy of the responses of halo-matter and halo-auto power spectra to $h$ calculated by {\tt Dark Emulator}. 
The emulator can predict the response accurately down to the nonlinear scale. 
For the halo-auto power spectrum, it turned out that {\tt Dark Emulator} does not predict the $h$ response correctly, and hence we also show the response to $A_s$ and $z$ calculated by {\tt Dark Emulator}, $T_{\ln A_s}^{\rm hh, emu}(k)$ and $T_{z}^{\rm hh, emu}(k)$. These responses approximate the response to $h$ well.  
Since the halo-auto power spectrum only contains the 2-halo term, it is less affected by a change in the concentration of halos and we can expect that the responses to $A_s$, $z$, and $h$ have similar features.

\begin{figure*}
	\includegraphics[width=\columnwidth]{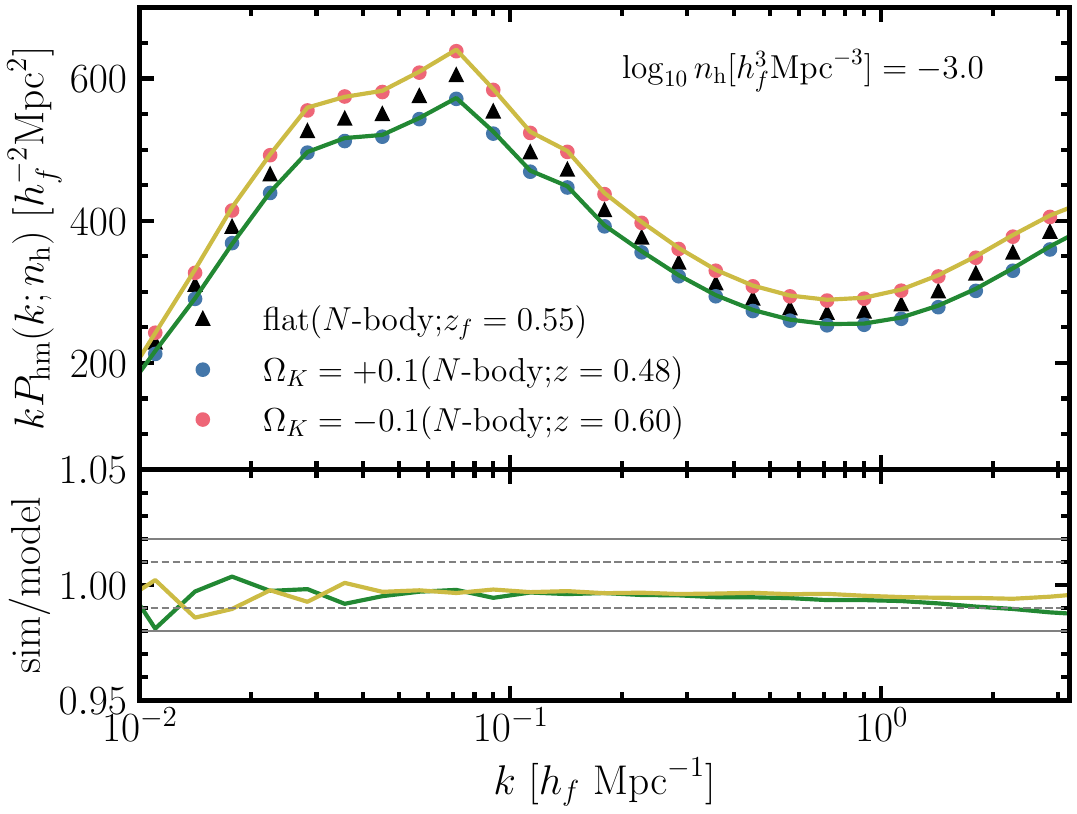}
    \includegraphics[width=\columnwidth]{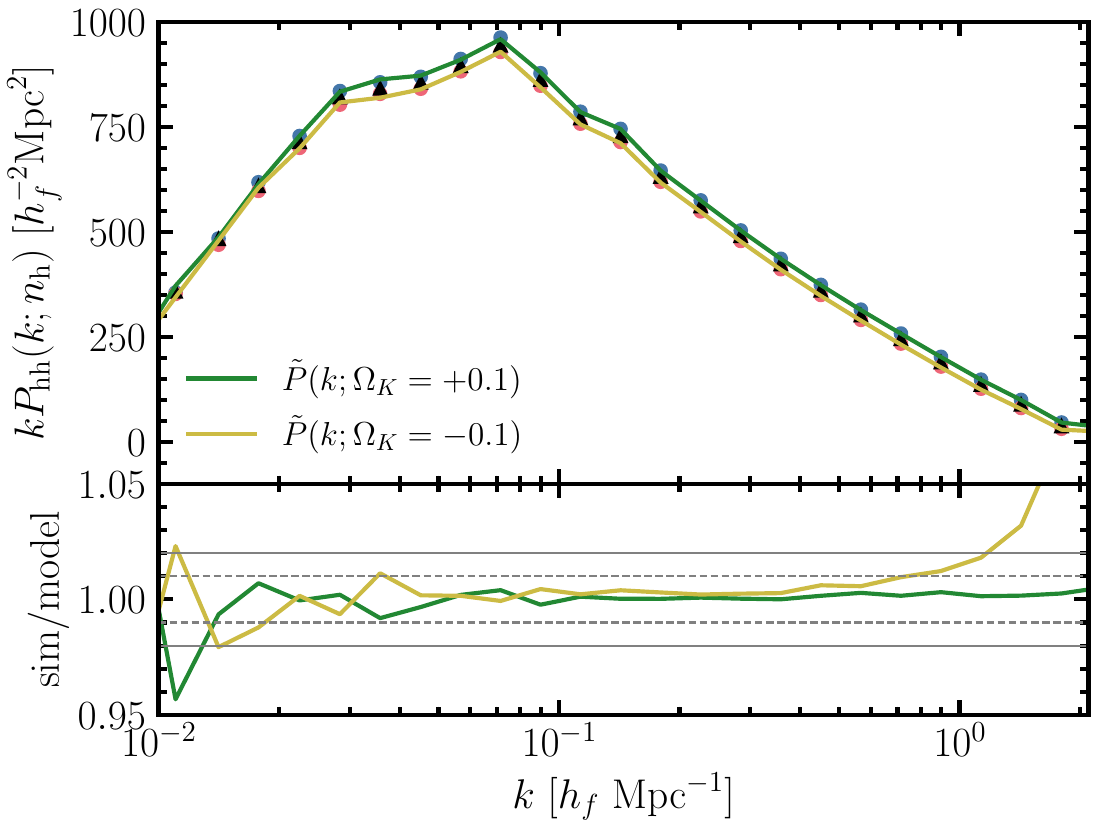}
\caption{An assessment of the accuracy of our method (Eq.~\ref{eq:pk_estimator}) for predicting the halo-matter power spectrum,
    $P_{\rm hm}(k)$ ({\em left panel}), and the halo-halo power spectrum, $P_{\rm hh}(k)$ ({\em right}), for nonflat $\Lambda$CDM models. The different symbols in each panel denote $P(k)$,
    directly estimated from $N$-body simulations for nonflat 
    models with $\Omega_K= \pm0.1$ ($\KCDM{2}$ models in Table~\ref{tab:simulations}), at $z_f =0.55$, while the lines denote the results from our method, $\tilde{P}(k)$ in Eq.~(\ref{eq:pk_estimator})
    Note that we used the simulation results for $P^f(k,z_f)$ and $T_h(k,z_f)$ in Eq.~(\ref{eq:pk_estimator}). 
    For the simulation results we used the ``paired-and-fixed''
    method in~\citet{2016MNRAS.462L...1A} to reduce the stochasticity, 
    and we considered the halo sample with number density $n_{\rm h} = 10^{-3}~h_f^{3}~{\rm Mpc}^{-3}$.
    We only plot the range of scales where the halo power spectra after subtracting the shot noise have positive values.
    For comparison, we also show the simulation result for the flat fiducial simulation by triangle symbols. 
    The lower plot in each panel shows the ratio between the simulation result and our method. The horizontal solid and dashed lines denote $\pm 2, \pm 1\%$ fractional accuracy, respectively. 
        }
	\label{fig:halo-matter-Okpm01}
\end{figure*}
\begin{figure*}
	\includegraphics[width=\columnwidth]{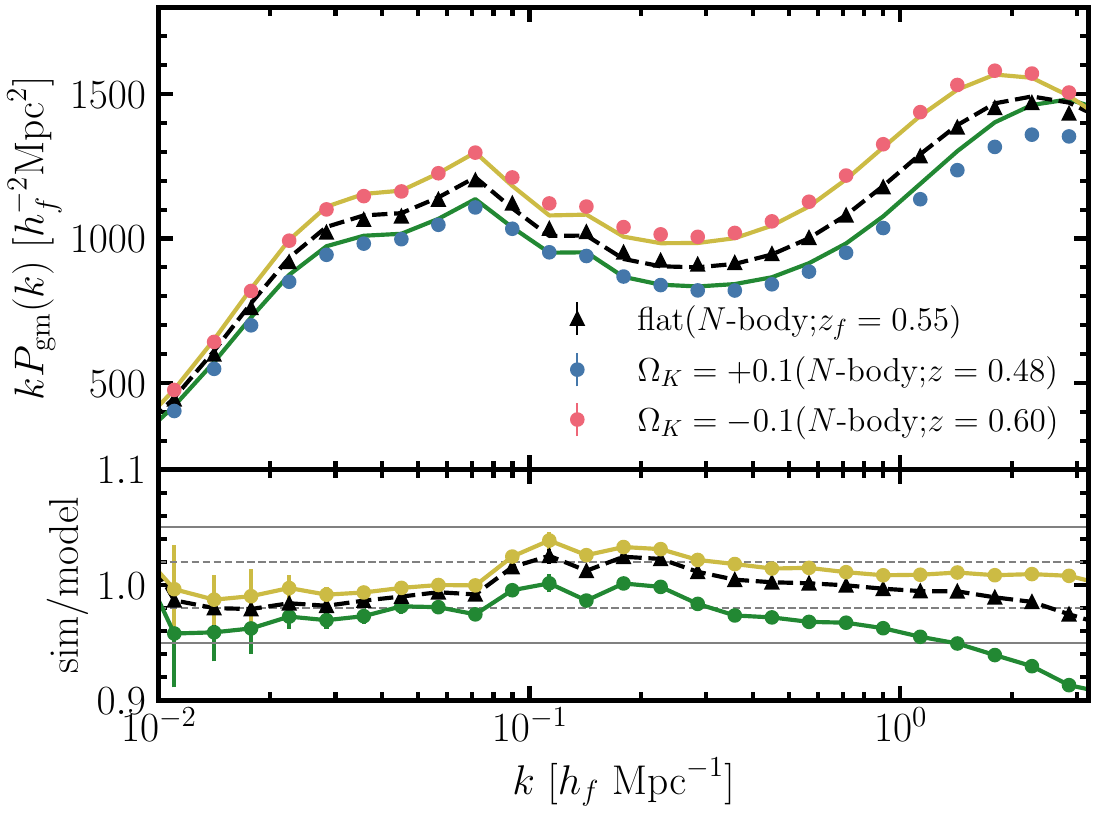}
	\includegraphics[width=\columnwidth]{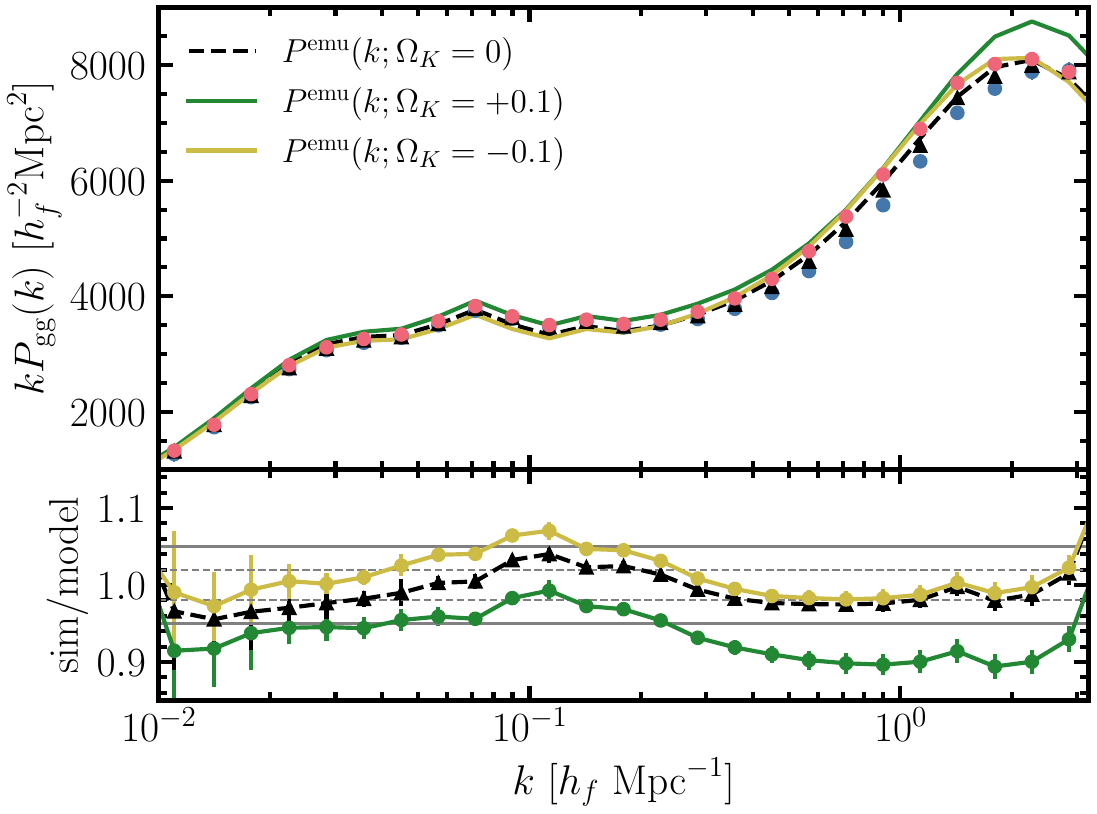}
\caption{An assessment of the accuracy of our method (Eq.~\ref{eq:pk_estimator}) for predicting 
the galaxy-matter power spectrum,
    $P_{\rm gm}(k)$ ({\em left panel}), and the galaxy-auto power spectrum, $P_{\rm gg}(k)$ ({\em right}), for nonflat $\Lambda$CDM models. The different symbols in each panel denote $P(k)$,
    directly estimated from $N$-body simulations for flat and nonflat 
    models with $\Omega_K= \pm0.1$, at $z_f =0.55$, while the lines denote the prediction using our method, $\tilde{P}(k)$ in Eq.~(\ref{eq:pk_estimator})
    together with {\tt Dark Emulator}. Error bars are estimated from the standard deviations among the 20 mock catalogs.
    The lower plot in each panel shows the ratio between the simulation result and our method. The horizontal solid and dashed lines denote $\pm 5, \pm 2\%$ fractional accuracy, respectively. 
        }      
	\label{fig:galaxy-matter-Okpm01}
\end{figure*}

\subsection{SU approach for $P(k;\Omega_K)$}

To assess the performance of our estimator for the halo-matter and halo-auto power spectra for nonflat universe $\tilde{P}(k;\Omega_K)$ (Eq.~\ref{eq:pk_estimator}), we compare it with the power spectra measured from $N$-body simulations with $\Omega_K = \pm 0.1$ (\KCDM2 model). 

In Fig.~\ref{fig:halo-matter-Okpm01}, the data points show the power spectrum measured from simulations, 
for \fCDM~model and \KCDM2 models. The curves show the predictions of Eq.~\ref{eq:pk_estimator}, 
where we used $P^f(k,z_f)$ and $T_h(k,z_f)$ measured from simulations of \fCDM~and $h$-\LCDM~models, respectively.
For the halo-matter power spectrum (left panel), the estimator has $\sim 1\%$ accuracy even for such large curvature, $\Omega_K = +0.1, -0.1$ (corresponding to $\deltab(z_f=0.55) = -0.12, +0.10$).
For the halo-auto power spectrum (right panel), the estimator has $\sim 2\%$ accuracy up to $k \simeq 1~h_f \rm{Mpc}^{-1}$. The relatively large deviation at $k \gtrsim 1~h_f \rm{Mpc}^{-1}$ is due to both large deviation from  $P^f(k,z_f)$ and the inaccuracy of approximation, $T_{\deltab}(k) \approx T_h(k)$.
The smaller fractional change in the amplitudes of $P_{\rm hm}$ and $P_{\rm hh}$
by the non-zero $\Omega_K$ than the change in $P_{\rm mm}$ \citep[see Fig.~5 in Ref.][]{Terasawa+22} is ascribed to the result of Fig.~\ref{fig:Tmm-hm-hh} (see the discussion around the figure).

Further, we tested the accuracy of the prediction for the galaxy-matter and galaxy-auto power spectra by {\tt Dark Emulator}, using the estimator in Eq.~(\ref{eq:pk_estimator}). 
We calculated these spectra as the weighted integral of $P_{\rm hm}(k;>M)$ or $P_{\rm hh}(k;>M)$ as in Eq.~\ref{eq:Pgm}. 
For the nonflat cosmology, we use the Eq.~\ref{eq:pk_estimator} to estimate 
$P_{\rm hm}(k;>M)$ and $P_{\rm hh}(k;>M)$, using $P^f(k,z_f)$ and $T_h(k,z_f)$ predicted by the emulator.
We used the halo mass function predicted by the emulator for the flat model, while we used the model in Ref~\citep{Tinker08} for the nonflat model because the emulator is not trained for the nonflat model.
On the large scale, we stitched the predictions with those of the linear theory, similarly as Eq.~\ref{eq:stitch}.

In Fig.~\ref{fig:galaxy-matter-Okpm01}, we can see the predictions of the galaxy-matter power spectrum by the emulator, $P^{\rm emu}(k)$ have $\sim 5 \%$ accuracy. Compared to the prediction for the flat universe, we can extend the prediction to the non-zero curvature without significant degradation.

For the galaxy-auto power spectrum, due to the poor accuracy of calculating the $h$ response by the emulator (Fig.~\ref{fig:halo-matteer-DE-Tbh-nth}), we use $T_{\ln A_s}^{\rm hh}(k)$ instead. The accuracy of predicting the galaxy-auto power spectra in nonflat universe is degraded compared to that for the flat universe, but the estimator still can predict the spectra with $\sim 10 \%$ accuracy.

\begin{figure}
	\includegraphics[width=\columnwidth]{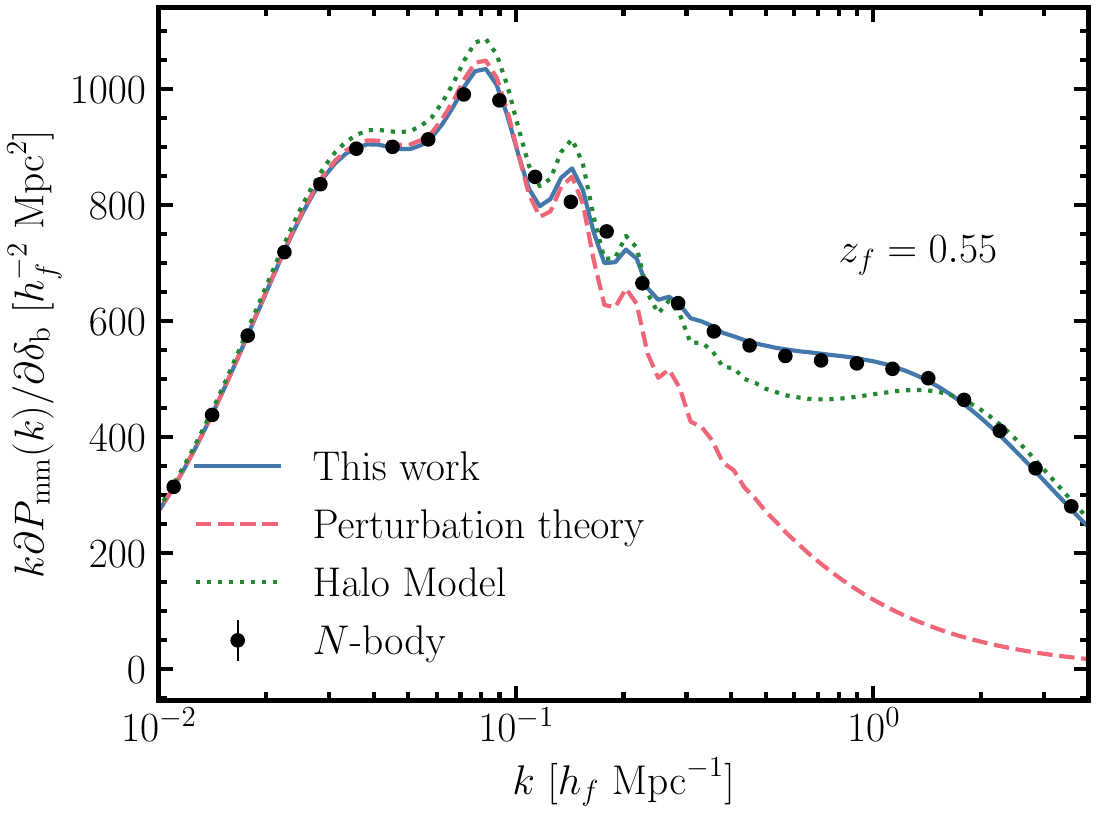}
\caption{Total response of matter power spectrum at $z_f = 0.55$ measured from the simulations and those calculated using our method (Eq.~\ref{eq:p_mm_total}; labeled as ``This work"), perturbation theory (Eq.~\ref{eq:B16response}) and halo model (Eq.~\ref{eq:HMresponse}). The error bars (although not visible) are estimated from the standard deviations among the 8 paired simulations.}
	\label{fig:Pmm_total_resp}
\end{figure}
\begin{figure*}
	\includegraphics[width=\columnwidth]{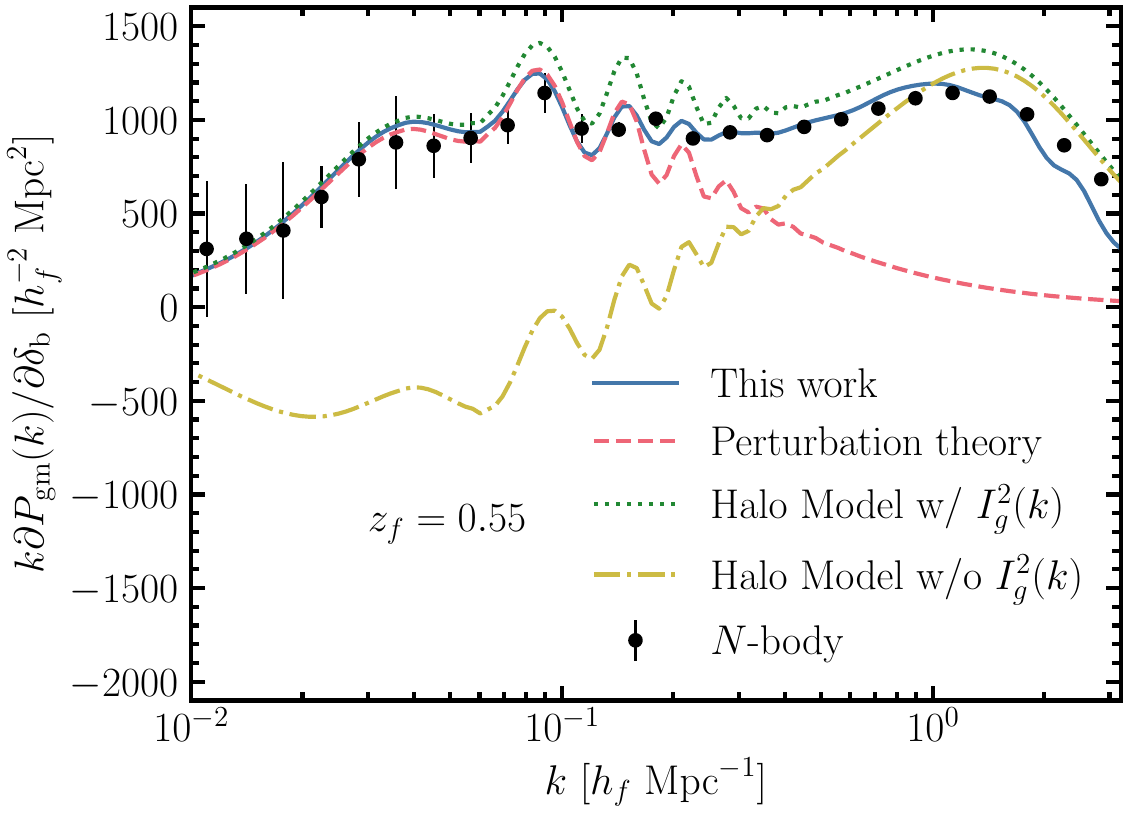}
	\includegraphics[width=\columnwidth]{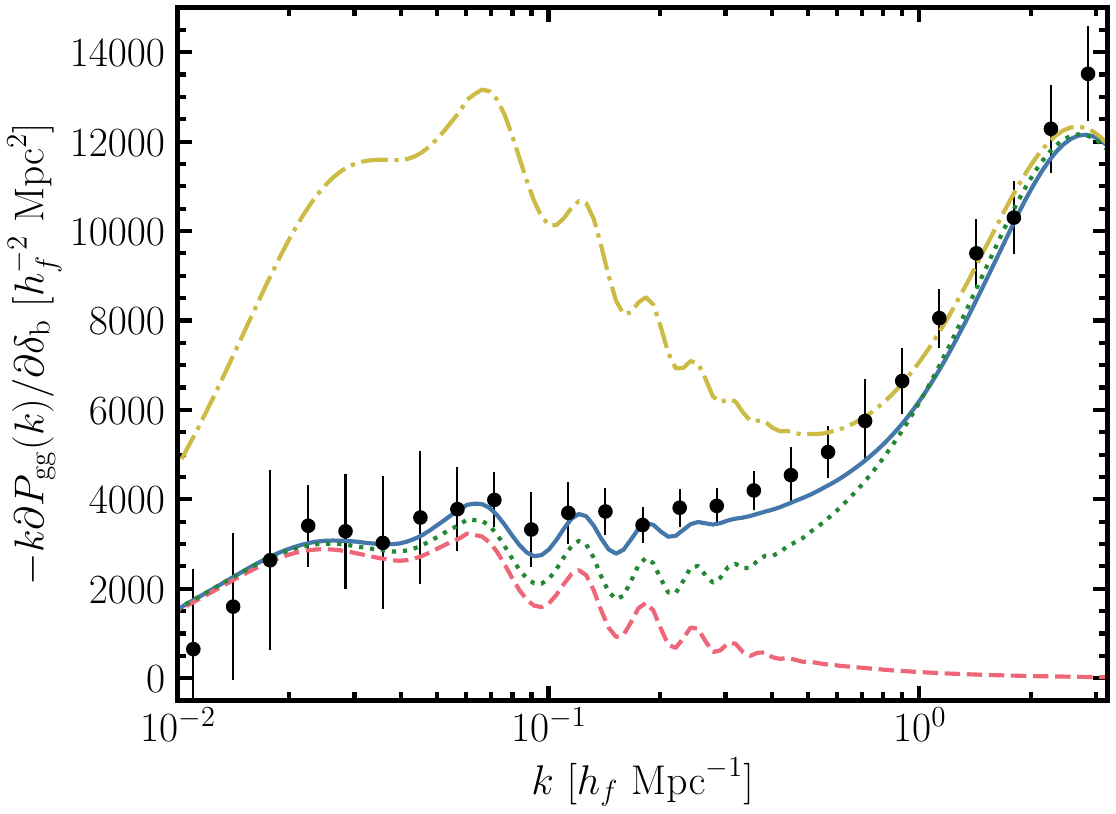}
\caption{Total response of power spectrum at $z_f = 0.55$ measured from the simulations and those calculated using our method (Eqs.~\ref{eq:p_XY_total}, \ref{eq:G^gm}, and \ref{eq:G^gg}); labeled as ``This work"), perturbation theory (Eq.~\ref{eq:B16response}) and halo model. The dotted and dashed-dotted lines denote the halo model (+HOD) prediction with or without $I_g^2(k)$ term
in Eq.~(\ref{eq:HMresponse}). Left: the response of galaxy-matter power spectrum. Right: the response of galaxy-auto power spectrum. The error bars are estimated from the standard deviations among the 8 paired simulations.
}
	\label{fig:PgX_total_resp}
\end{figure*}

\subsection{Total response}

We tested the accuracy of our estimator for the power spectrum total response to the super-survey modes compared with the measurement from the mock galaxy catalog and $N$-body results.

In Fig.~\ref{fig:Pmm_total_resp} we show the accuracy of the matter power spectrum total-response at $z_f = 0.55$ calculated by our method (Eq.~\ref{eq:p_mm_total}) against the $N$-body simulations, along with other theoretical predictions. We use {\tt Halofit} in Smith et al. (Smith+03)~\citep{Smith03} and Takahashi et al. (Takahashi+12)~\citep{Takahashi12} to compute $T_h^{\rm mm}(k)$ and $P_{\rm mm}(k)$ in Eq.~(\ref{eq:p_mm_total}), respectively. This different choice of {\tt Halofit} is because Smith+03 predicts $T_h^{\rm mm}(k)$ better than Takahashi+12 (See the Fig.~3 in~\citep{Terasawa+22}), while Takahashi+12 is known to be more accurate for $P_{\rm mm}(k)$ especially on small scales. 
The perturbation theory prediction agrees with the $N$-body results 
on large scales, but 
it begins to deviate around $k \gtrsim 0.1 \, h_f {\rm Mpc}^{-1}$. The halo model reproduces the behavior of the response out to the nonlinear scale, but 
it underestimates the response by $\sim 20 \%$ at $k \sim 1 \, h {\rm Mpc}^{-1}$.
Compared with these two analytical models which are commonly used in the literature, our method predicts the response better over a wide range of scales. 

In Fig.~\ref{fig:PgX_total_resp} we show the accuracy of the galaxy-matter and galaxy-auto power spectrum total response at $z_f = 0.55$ calculated by our method (Eqs.~\ref{eq:p_XY_total}, \ref{eq:G^gm}, and \ref{eq:G^gg}.) against the $N$-body simulations, along with other theoretical predictions. We use {\tt Dark Emulator} to compute $P(k)$, $T_h(k)$, and mass function in Eqs.~(\ref{eq:p_XY_total}), (\ref{eq:G^gm}), and (\ref{eq:G^gg}).
For the galaxy-auto power spectrum, as discussed in the Sec.~\ref{subsec:Growth response}, since {\tt Dark Emulator} does not predict the $h$ response of the halo-auto power spectrum correctly, we used the response to $A_s$ calculated by {\tt Dark Emulator} instead. 
As with the matter power spectrum response, our method can predict the galaxy-matter and galaxy-auto power spectrum responses more accurately compared to the analytical ones in the literature. 

Lastly, we mention the linear limit of the halo model predictions. In Figs.~\ref{fig:Pmm_total_resp} and \ref{fig:PgX_total_resp}, as we discussed in Sec.~\ref{sec:Preliminary}, we can see the predictions of the halo model converge to those of the perturbation theory at large scale. 
We also show the impact of ignoring the $I_{\rm g}^2(k)$ term in the halo model prediction. It is clear that ignoring the term results in an inaccurate prediction on linear and quasi-nonlinear scales.

\section{conclusion}\label{sec:conclusion}

In this paper, we have developed an approximate method to model the halo-matter and halo-auto power spectra for the nonflat \LCDM~model, from quantities representing the nonlinear evolution of the corresponding flat \LCDM~model, based on the SU method.
The key points to build the estimator are the correspondence between the nonflat and flat universes through the SU picture and the equivalence of the growth responses to long-wavelength modes and the Hubble parameter.
This work is a sequel of our previous research~\cite{Terasawa+22}, in which we proposed the approximate method for predicting the nonlinear matter power spectrum for nonflat \LCDM~model.

The estimator of the halo-matter (halo-auto) power spectrum has $\sim 1\%$ ($\sim 2\%$) accuracy even for large curvature model with $\Omega_K = \pm 0.1$.
Using the estimator we can extend the existing emulators to predict the nonlinear power spectra for
nonflat universe without degrading its accuracy. In particular, we showed we can extend the {\tt Dark Emulator} to predict the galaxy-matter and galaxy-auto power spectra for a
non-zero curvature model.

The response $T_{\deltab}(k)$ is also a key quantity for estimating SSC. We utilized the approximate identity $T_{\deltab}(k) \approx T_h(k)$ and proposed the calculation of the total response (SSC terms) using
{\tt Halofit} or {\tt Dark Emulator}. 
We showed that our method can predict the total response at an accuracy
better than the analytical methods used in the literature such as the perturbation theory and the halo model, thanks to capturing the nonlinear response through $T_h(k)$.
Our method for computing the total response would be implemented into the Core Cosmology Library~\citep{CCL}.

Although we assume \LCDM~model in this paper, we expect our approach to approximate the response to $\deltab$ is also applicable to $w$CDM model, where dark energy is not the cosmological constant. Once validated with numerical simulations, we can ease the computational cost of the SU simulations by substituting it with the response to another parameter which can be evaluated fast. 
In this paper, we focused on the responses to $\deltab$ and $h$, but we can think of other parameters that share the almost same response in wider cosmological parameter space, which can be used to extend the simulation-based theory predictions and to ease the computational cost for building such models.

\acknowledgments
We would like to thank Kaz~Akitsu, Yosuke~Kobayashi, Yue~Nan, Tim~Eifler, and Elisabeth~Krause for their useful and stimulating discussion. 
This work was supported in part by World Premier International Research Center Initiative (WPI Initiative), MEXT, Japan, JSPS KAKENHI Grant Numbers JP23KJ0747, JP22H00130, JP20H05850, JP20H05855, JP20H05861, JP20H04723, JP19H00677, JP21H01081, JP22K03634, Basic Research Grant (Super AI) of Institute for AI and Beyond of the University of Tokyo, and Japan Science and Technology Agency (JST) AIP Acceleration Research Grant Number JP20317829.
Numerical computations were carried out on Cray XC50 at Center for Computational Astrophysics, National Astronomical Observatory of Japan.

\onecolumngrid
\appendix

\section{Calculation methods of the response functions of halo 
and galaxy power spectra}
\label{sec:derivation}

\subsection{The response of galaxy-matter power spectrum}
\label{subsec:Pgm}
We define the cumulative number density of halos as
\begin{align}
	n(>M_{\rm th};\deltab) \equiv \int_{\rm M_{th}}^{\infty} \frac{\mathrm{d}M}{M} n_{\ln M}(M;\deltab),
\end{align}
where $n_{\ln M} \equiv \mathrm{d}n/\mathrm{d}\ln M$.
We define the following dimensionless quantities:
\begin{eqnarray}
\mathscr{P}_\mathrm{hm}(k;>M_\mathrm{th}) \equiv n(>M_\mathrm{th})P_\mathrm{hm}(k;>M_\mathrm{th}),
\end{eqnarray}
for mass-threshold samples (above mass $M_\mathrm{th}$) and
\begin{eqnarray}
\mathscr{P}_\mathrm{hm}(k;M_-,M_+) \equiv n(M_-,M_+)P_\mathrm{hm}(k;M_-,M_+),
\end{eqnarray}
for mass bin samples in the mass range of $[M_-,M_+]$, respectively.
Using these quantities, all the equations for $P_\mathrm{gm}$ and its response can be 
simplified. The idea behind this is as follows. First, mass-threshold halo samples are easier to analyze in simulations. Once the halo-matter power spectrum for various different mass thresholds is known, one can convert them to that for mass-bin samples by taking a derivative w.r.t. the mass threshold. Numerically, this can be done by taking the finite difference:
\begin{equation}
P_\mathrm{hm}(k;M) \simeq
\dfrac{n(>M_-) P_\mathrm{hm}(k;>M_-) - n(>M_+) P_\mathrm{hm}(k;>M_+)}{n(>M_-) - n(>M_+)}, 
\label{eq:Phm_at_M}
\end{equation}
where $M_\pm = M \pm \epsilon_M$ with some small 
$\epsilon_M$ compared to $M_\pm$.
In the above, note that the halo-matter power spectrum is a halo number-weighted quantity, and thus $n$ must be considered appropriately. This equation is actually implemented in {\tt Dark Emulator}. Now, using $\mathscr{P}_\mathrm{hm}$, this relation is simplified as
\begin{eqnarray}
\mathscr{P}_\mathrm{hm}(k;M_-,M_+) =  \mathscr{P}_\mathrm{hm}(k;>M_-) - \mathscr{P}_\mathrm{hm}(k;>M_+). 
\end{eqnarray}
We can use a similar trick for the galaxy-matter power spectrum. Assuming a halo occupation distribution model, we can have
\begin{eqnarray}
P_\mathrm{gm}(k)=\frac{1}{\bar{n}_\mathrm{g}}
\int\mathrm{d} M \langle \tilde{U}_g(k;M)\rangle
\dfrac{\mathrm{d} n}{\mathrm{d}M}(M)P_\mathrm{hm}(k;M), 
\label{eq:Pgm}
\end{eqnarray}
where $\tilde{U}_g (k;M)$ describes the Fourier transform of the average radial profile of galaxy number density 
in host halos
with mass $M$, which can be computed by taking the product of
the halo occupation
distribution $N_\mathrm{g}(M)$ and the radial profile of each galaxy.

This can be evaluated by substituting Eq.~(\ref{eq:Phm_at_M}). Now, multiplying both sides of Eq.~(\ref{eq:Pgm}) by $\bar{n}_\mathrm{g}$ and defining $\mathscr{P}_\mathrm{gm} = \bar{n}_\mathrm{g} P_\mathrm{gm}$, one can compute
\begin{eqnarray}
\mathscr{P}_\mathrm{gm}(k)
&=&
\int \mathrm{d}M \langle \tilde{U}_g(k;M)\rangle
\dfrac{\mathrm{d}}{\mathrm{d}M}\Bigl(
-\mathscr{P}_\mathrm{hm}(k;>M)
\Bigr)\nonumber\\
&=&
-\langle \tilde{U}_g(k;M)\rangle\mathscr{P}_\mathrm{hm}(k;>M)\Bigr|_{0}^{\infty} 
+\int \mathrm{d}M \dfrac{\mathrm{d}\langle \tilde{U}_g(k;M)\rangle}{\mathrm{d}M}\mathscr{P}_\mathrm{hm}(k;>M),
\nonumber\\
&=&
\int \mathrm{d}M \dfrac{\mathrm{d}\langle \tilde{U}_g(k;M)\rangle}{\mathrm{d}M}\mathscr{P}_\mathrm{hm}(k;>M). 
\end{eqnarray}
In the above, we have performed an integration by parts to obtain the second line and then use the fact that $N_\mathrm{g}(M)$ tends to zero at the low mass end and $\mathscr{P}_\mathrm{hm}(M)$ tends to zero at the high mass end to reach the final line.
This is how we can avoid estimating the power spectrum for the mass-bin samples and directly compute the galaxy-matter power spectrum from the mass-threshold halo samples.

The responses are also related by a similar equation as
\begin{eqnarray}
\left. \frac{\partial \mathscr{P}_\mathrm{gm}(k)}{\partial \delta_\mathrm{b}} \right|_{\rm G}
=  
\int \mathrm{d}M \left[ \frac{\mathrm{d}\langle \tilde{U}_g(k;M)\rangle}{\mathrm{d}M} 
\left.\frac{\partial\mathscr{P}_\mathrm{hm}(k;>M)}{\partial \delta_\mathrm{b}}\right|_{{\rm G}, M} +\frac{\partial}{\partial \delta_\mathrm{b}} \left. \frac{\mathrm{d}\langle \tilde{U}_g(k;M)\rangle}{\mathrm{d}M} \right|_{\rm G}  \mathscr{P}_\mathrm{hm}(k;>M) \right],
\end{eqnarray}
where the subscript $M$ denotes the derivative with fixed mass threshold. Assuming that the cosmological dependence is fully encoded in $\mathscr{P}_\mathrm{gm}$ and that the galaxy profile around halos 
in physical scale and the mean HOD are given irrespective of the cosmological model, which is partly validated in Ref~\citep{2021JCAP...05..069V}, one can trivially derive 
\begin{eqnarray}
\frac{\partial}{\partial \delta_\mathrm{b}} \left. \frac{\mathrm{d}\langle \tilde{U}_g(k;M)\rangle}{\mathrm{d}M} \right|_{\rm G} = 
\frac{1}{3}
\frac{\partial}{\partial \ln k} \frac{\mathrm{d}\langle \tilde{U}_g(k;M)\rangle}{\mathrm{d}M}.
\end{eqnarray}

Finally, one can convert this to the standard response function by
\begin{eqnarray}
\left. \dfrac{\partial P_\mathrm{gm}(k)}{\partial \delta_\mathrm{b}}\right|_{\rm G} = \frac{1}{\bar{n}_\mathrm{g}} \left.\dfrac{\partial \mathscr{P}_\mathrm{gm}(k)}{\partial \delta_\mathrm{b}}\right|_{\rm G} - b_\mathrm{g}^\mathrm{L} P_\mathrm{gm}(k),
\label{eq:Pgm_growth}
\end{eqnarray}
where $b_g^L$ is the Lagrangian galaxy bias, related to the Eulerian bias as $b_g = b_g^L + 1$. 
We can give a physical interpretation of the bias factor as follows. The Lagrangian galaxy bias, $b_g^L$, describes the change in the number density of galaxies at a fixed Lagrangian, or, equivalently, \textit{local} comoving 
volume in the SU picture, and the difference between the Eulerian and the Lagrangian bias, $b_g - b_g^L = 1$  represents the contribution from the change in the physical volume, i.e., equivalent to the dilation effect. Note that the second term in Eq.~\ref{eq:Pgm_growth}, $-b_g^L P_{\rm gm}(k)$ comes from the response of $\bar{n}_g$ and since the growth response should be evaluated at a fixed comoving volume, the response of $\bar{n}_g$ gives $b_g^L$ instead of $b_g$.

\subsection{The response of galaxy auto-power spectrum}
Similarly, we can also evaluate the galaxy-auto power spectrum using mass-threshold quantities. 
First, the halo-auto power spectrum of the mass-bin samples can be calculated by taking the finite difference:
\begin{align}
  P_{\rm hh}(k;M_1,M_2) = \frac{\frac{\partial^2}{\partial M \partial M'}
  \left.
  \left[n(>M) n(>M') P_{\rm hh}(k;>M, >M') \right]\right|_{M=M_1,M'=M_2}}
  {\frac{\mathrm{d}n}{\mathrm{d}M}(M_1) \frac{\mathrm{d}n}{\mathrm{d}M}(M_2)}.
  \label{eq:phh-massbin}
\end{align}
Following the halo model description, we split the galaxy power spectrum into the 2-halo and 1-halo terms as $P_{\rm gg}(k) = P_{\rm gg}^{2h}(k) + P_{\rm gg}^{1h}(k)$.
Using Eq.~\ref{eq:phh-massbin}, we can write the 2-halo term of the galaxy power spectrum as
\begin{align}
  P_{\rm gg}^{2h}(k) &= \frac{1}{\bar{n}_g^2} \int \mathrm{d}M \frac{\mathrm{d}n}{\mathrm{d}M}(M) \langle \tilde{U}_g (k;M)\rangle \int \mathrm{d}M' \frac{\mathrm{d}n}{\mathrm{d}M'}(M')  \langle\tilde{U}_g(k;M')\rangle P_{\rm hh}(k; M,M') \nonumber \\
  &= \frac{1}{\bar{n}_g^2} \int \mathrm{d}M \frac{\mathrm{d}\langle \tilde{U}_g (k;M)\rangle}{\mathrm{d}M} n(>M) \int \mathrm{d}M' \frac{\mathrm{d} \langle\tilde{U}_g(k;M')\rangle}{\mathrm{d}M'} n(>M') P_{\rm hh}(k; >M, >M') \nonumber \\
  &=\frac{1}{\bar{n}_g^2} \iint \mathrm{d}M \mathrm{d}M' \frac{\mathrm{d}\langle \tilde{U}_g (k;M)\rangle}{\mathrm{d}M}  \frac{\mathrm{d}\tilde{U}_g(k;M')}{\mathrm{d}M'} 
  \mathscr{P}_{\rm hh}(k;>M,>M'),
\end{align}
where we did integral by parts and ignored the surface terms, and 
we define 
\begin{align}
  \mathscr{P}_{\rm hh}(k;>M_{\rm th},>M'_{\rm th})  
  &\equiv n(>M_{\rm th}) n(>M'_{\rm th}) 
  P_{\rm hh}(k;>M_{\rm th},>M'_{\rm th}).
\end{align}

The SU growth responses of the 2-halo term of the galaxy-auto power spectrum can be calculated as
\begin{align}
  \left.\parfrac{P_{\rm gg}^{2h}(k)}{\deltab}\right|_{\rm G} 
  &=\frac{1}{\bar{n}_g^2} \iint \mathrm{d}M \mathrm{d}M' \frac{\mathrm{d}\tilde{U}_g(k;M')}{\mathrm{d}M'} \left[
  \frac{\mathrm{d}\langle \tilde{U}_g (k;M)\rangle}{\mathrm{d}M}  
  \left.\parfrac{\mathscr{P}_{\rm hh}(k;>M,>M')}{\deltab}\right|_{{\rm G}, M, M'} \right. \nonumber \\ 
  &~~~~~ + \left. \left. 2 \frac{\partial}{\partial\deltab}  \frac{\mathrm{d}\langle \tilde{U}_g (k;M)\rangle}{\mathrm{d}M} \right|_{\rm G}  
  {\mathscr{P}_{\rm hh}(k;>M,>M')} \right]
  - 2b_g^L P_{\rm gg}^{2h}(k).
\end{align}

For the 1-halo term, we can compute the SU growth response as
\begin{align}
  \left.\parfrac{P_{\rm gg}^{1h}(k)}{\deltab}\right|_{\rm G} &=   
  \frac{1}{\bar{n}_g^2} \int \mathrm{d}M
  \frac{\mathrm{d}n}{\mathrm{d}M}(M)  \left[ b_{\mathrm{h}, 1}^L(M) \langle\tilde{U}_g^2(k;M)\rangle + \left.\frac{\partial \langle\tilde{U}_g^2(k;M)\rangle}{\partial \deltab} \right|_{\rm G}
  \right]
  - 2b_g^L P_{\rm gg}^{1h}(k). 
  \label{eq:G^gg1h}
\end{align}
Note that the response of {the} 1-halo term {in our approach} is identical to that of the halo model description~\cite[e.g.][]{cosmolike, 2020JCAP...03..044N}.

\subsection{The response calibration of halo and galaxy power spectra
using the abundance matching method}
\label{appendix:pk_response}
In this subsection, we describe the relation between the SU growth response and the number density fixed or abundance matched response which we clarify later.
We denote the derivative performed keeping the comoving halo number density fixed
as ``AM" response, which stands for Abundance Matching method.
When evaluating the AM response, 
we change the threshold $M_{\rm th}(\deltab)$ so that the cumulative halo number density in the comoving volume is kept fixed when varying $\deltab$~\citep{2016PhRvD..93f3507L}:
\begin{align}
	\frac{\mathrm{d}n(>M_{\rm th};\deltab)}{\mathrm{d}\deltab} = 0.
	\label{eq:dndb}
\end{align}
We define the mass threshold shift $s(M)$ to keep the number density as
\begin{align}
	\mathrm{d} \ln M_{\rm th} \equiv s(M_{\rm th}) \mathrm{d}\deltab.
\end{align}
We can calculate Eq.~(\ref{eq:dndb}) as
\begin{align}
	\frac{\mathrm{d}n(>M_{\rm th};\deltab)}{\mathrm{d}\deltab} &= \frac{\mathrm{d}}{\mathrm{d}\deltab} \int_{\rm M_{th}(\deltab)}^{\infty} \frac{\mathrm{d}M}{M} n_{\ln M}(M;\deltab) \nonumber \\
	&= \int_{\rm M_{th}}^{\infty} \frac{\mathrm{d}M}{M} \parfrac{n_{\ln M}(M;\deltab)}{\deltab} 
    + \frac{\partial}{\partial\deltab} \left[F_1(M=\infty) - F_1(M=M_{\rm th})  \right] \nonumber \\
	&= \int_{\rm M_{th}}^{\infty} \frac{\mathrm{d}M}{M} \parfrac{n_{\ln M}(M;\deltab)}{\deltab} 
    - n_{\ln M}(M_{\rm th}) s(M_{\rm th})  \nonumber \\
    &= 0,
	\label{eq:dndb_chain}
\end{align}
where we defined $F_1(M)$ as the function satisfying $\partial F_1(M)/\partial\ln M = n_{\ln M}(M)$
and we can calculate its response as    
\begin{align}
	\parfrac{F_1(M)}{\deltab} &= \parfrac{F_1(M)}{\ln M} \parfrac{\ln M}{\deltab} \nonumber \\
	&= n_{\ln M}(M) s(M).
\end{align}
Hence, we have
\begin{align}
	\int_{\rm M_{th}}^{\infty} \frac{\mathrm{d}M}{M} \parfrac{n_{\ln M}(M;\deltab)}{\deltab} = n_{\ln M}(M_{\rm th}) s(M_{\rm th}).
\end{align}
Using this, the Lagrangian halo bias above the mass threshold is given as~\citep{2016PhRvD..93f3507L}
\begin{align}
	b_{\mathrm{h}, 1}^L(>M_{\rm th}) &\equiv \frac{1}{n(>M_{\rm th})} \int_{M_{\rm th}}^{\infty} \frac{\mathrm{d}M}{M} b_{\mathrm{h}, 1}^L n_{\ln M} \nonumber \\
	&= \frac{1}{n(>M_{\rm th})} \int_{M_{\rm th}}^{\infty} \frac{\mathrm{d}M}{M} \parfrac{\ln n_{\ln M}}{\deltab} n_{\ln M} \nonumber \\
	&= \frac{n_{\ln M}(M_{\rm th})s(M_{\rm th})}{n(>M_{\rm th})}.
\end{align}
Likewise, the AM response of 
$\mathscr{P}_{\rm hm}(k;>M_{\rm th})$ can be calculated as
\begin{align}
	\left.\frac{\mathrm{d}\mathscr{P}_{\rm hm}(k;>M_{\rm th})}{\mathrm{d}\deltab}\right|_{\rm G} &= \frac{\mathrm{d}}{\mathrm{d}\deltab} \int_{\rm M_{th}}^{\infty} \mathrm{d}\ln M n_{\ln M}(M) P_{\rm hm}(k;M) \nonumber \\
	&=  \int_{\rm M_{th}}^{\infty} \mathrm{d}\ln M \parfrac{~}{\deltab} \left[n_{\ln M}(M) P_{\rm hm}(k;M) \right] 
    + \frac{\mathrm{d}}{\mathrm{d}\deltab} \left[F_2(M=\infty) - F_2(M=M_{\rm th})  \right]   \nonumber \\
	&= \left.\parfrac{\mathscr{P}_{\rm hm}(k;>M_{\rm th})}{\deltab}\right|_{{\rm G}, M_{\rm th}} 
    - n_{\ln M}(M_{\rm th}) P_{\rm hm}(k;M_{\rm th}) s(M_{\rm th}) \nonumber \\
	&= \left.\parfrac{\mathscr{P}_{\rm hm}(k;>M_{\rm th})}{\deltab}\right|_{{\rm G}, M_{\rm th}}  
    -  b_{\mathrm{h}, 1}^L(>M_{\rm th}) n(>M_{\rm th}) P_{\rm hm}(k;M_{\rm th}),
  \label{eq:dnPdb_nfix}
\end{align}
where $F_2(M)$ is a function satisfying $\partial F_2(M)/\partial\ln M = n_{\ln M}(M)P_{\rm hm}(k;M)$.
On the other hand, the l.h.s. of Eq.~(\ref{eq:dnPdb_nfix}) can be written as
\begin{align}
  \left.\frac{\mathrm{d}\mathscr{P}_{\rm hm}(k;>M_{\rm th})}{\mathrm{d}\deltab}\right|_{\rm G} &= n(>M_{\rm th}) \left.\parfrac{P_{\rm hm}(k;>M_{\rm th})}{\deltab}\right|_{\rm G} \nonumber \\
  &= n(>M_{\rm th}) \left(\frac{26}{21}\right) P_{\rm hm}(k;>M_{\rm th}) 
  \, T_{\deltab}^{\rm hm}(k;>M_{\rm th}) \nonumber \\
  &\simeq n(>M_{\rm th}) \left(\frac{26}{21}\right) P_{\rm hm}(k;>M_{\rm th}) 
  \, T_{h}^{\rm hm}(k;>M_{\rm th}),
\end{align}
where we used the approximate relation $T_{h}^{\rm hm}(k;>M_{\rm th}) \approx T_{\deltab}^{\rm hm}(k;>M_{\rm th})$ in the last equation.
Hence, we can calculate the
$P_{\rm gm}(k)$ SU-growth response using $T_{h}(k;>M_{\rm th})$ as
\begin{align}
  \left.\parfrac{P_{\rm gm}(k)}{\deltab}\right|_{\rm G} &= \frac{1}{\bar{n}_g}\left.\parfrac{\mathscr{P}_\mathrm{gm}(k)}{\deltab}\right|_{\rm G} - b_g^L P_{\rm gm}(k) \nonumber \\
  &= \frac{1}{\bar{n}_g}\int \mathrm{d}M \left\{ \dfrac{\mathrm{d}\langle \tilde{U}_g (k;M)\rangle}{\mathrm{d}M} \left.\parfrac{\mathscr{P}_\mathrm{hm}(k;>M)}{\deltab}\right|_{{\rm G}, M} + \frac{1}{3} \mathscr{P}_\mathrm{hm}(k;>M)
\frac{\partial}{\partial \ln k} \frac{\mathrm{d}\langle \tilde{U}_g(k;M)\rangle}{\mathrm{d}M} \right\}
  - b_g^L P_{\rm gm}(k)   \nonumber \\
  &= \frac{1}{\bar{n}_g}\int \mathrm{d}M \left\{ \dfrac{\mathrm{d}\langle \tilde{U}_g (k;M)\rangle}{\mathrm{d}M} n(>M) 
  \left[ b_{\mathrm{h}, 1}^L(>M)P_{\rm hm}(k;M) 
  +\left(\frac{26}{21}\right) T_{h}^{\rm hm}(k;>M) P_{\rm hm}(k;>M) \right] \right. \nonumber \\
  &~~~~~ + \left. \frac{1}{3} P_\mathrm{hm}(k;>M)
\frac{\partial}{\partial \ln k} \frac{\mathrm{d}\ln \langle \tilde{U}_g(k;M)\rangle}{\mathrm{d}M} \right\} - b_g^L P_{\rm gm}(k).
  \label{eq:Appendix_G^gm} 
\end{align}

The AM response of $\mathscr{P}_{\rm hh}(k;>M_{\rm th},>M'_{\rm th})$ is calculated as
\begin{align}
  \left.\frac{\mathrm{d}\mathscr{P}_{\rm hh}}{\mathrm{d}\deltab}\right|_{\rm G} &= \left.\frac{\mathrm{d}\mathscr{P}_{\rm hh}(k;>M_{\rm th},>M'_{\rm th})}{\mathrm{d}\deltab}\right|_{{\rm G}, M_{\rm th}, M'_{\rm th}}  
  + \int_{M_{\rm th}}\mathrm{d}\ln M n_{\ln M}(M) \left\{ 
  \frac{\partial}{\partial\deltab} \left[F_3(M'=\infty) - F_3(M'=M'_{\rm th}) \right] \right\} \nonumber \\
  &~~~~~+ \int_{M'_{\rm th}}\mathrm{d}\ln M' n_{\ln M}(M') \left\{ 
  \frac{\partial}{\partial\deltab} \left[F_3(M=\infty) - F_3(M=M_{\rm th}) \right] \right\} \nonumber \\
  &= \left.\frac{\mathrm{d}\mathscr{P}_{\rm hh}(k;>M_{\rm th},>M'_{\rm th})}{\mathrm{d}\deltab}\right|_{{\rm G}, M_{\rm th}, M'_{\rm th}} 
  - b_{\mathrm{h}, 1}^L(>M'_{\rm th}) n(>M'_{\rm th}) 
  \int_{M_{\rm th}}\mathrm{d}\ln M n_{\ln M}(M) P_{\rm hh}(k;M,M'_{\rm th}) \nonumber \\
  &~~~~~- b_{\mathrm{h}, 1}^L(>M_{\rm th}) n(>M_{\rm th}) 
  \int_{M'_{\rm th}}\mathrm{d}\ln M' n_{\ln M}(M') P_{\rm hh}(k;M',M_{\rm th}) \nonumber \\
  &= \left.\frac{\mathrm{d}\mathscr{P}_{\rm hh}(k;>M_{\rm th},>M'_{\rm th})}{\mathrm{d}\deltab}\right|_{{\rm G}, M_{\rm th}, M'_{\rm th}} 
  - n(>M_{\rm th})n(>M'_{\rm th}) \nonumber \\
  &~~~~~\times \left[ b_{\mathrm{h}, 1}^L(>M'_{\rm th}) P_{\rm hh}(k;>M_{\rm th},M'_{\rm th}) 
  + b_{\mathrm{h}, 1}^L(>M_{\rm th}) P_{\rm hh}(k;>M'_{\rm th},M_{\rm th}) \right],
  \label{eq:dMdb_nfix}
\end{align}
where $F_3(M)$ satisfying 
$\partial F_3(M)/\partial\ln M = n_{\ln M}(M)P_{\rm hh}(k;M,M')$. 
Since the l.h.s. of Eq.~(\ref{eq:dMdb_nfix}) can be written using the approximate relation $T_{\deltab}^{\rm hh}(k;>M_{\rm th},>M'_{\rm th}) \approx T_{h}^{\rm hh}(k;>M_{\rm th},>M'_{\rm th})$ as 
\begin{align}
  \left.\frac{\mathrm{d}\mathscr{P}_{\rm hh}}{\mathrm{d}\deltab}\right|_{\rm G} &= n(>M_{\rm th}) n(>M'_{\rm th}) \left.\frac{\mathrm{d}P_{\rm hh}(k;>M_{\rm th},>M'_{\rm th})}{\mathrm{d}\deltab}\right|_{\rm G}    \nonumber \\
  &= n(>M_{\rm th}) n(>M'_{\rm th}) \left(\frac{26}{21}\right) 
  T_{\deltab}^{\rm hh}(k;>M_{\rm th},>M'_{\rm th}) 
  P_{\rm hh}(k;>M_{\rm th},>M'_{\rm th})\nonumber \\
  &\simeq n(>M_{\rm th}) n(>M'_{\rm th}) \left(\frac{26}{21}\right) 
  T_{h}^{\rm hh}(k;>M_{\rm th},>M'_{\rm th}) 
  P_{\rm hh}(k;>M_{\rm th},>M'_{\rm th}).
\end{align}
The SU-growth response of 
$P_{\rm gg}^{2h}$ can be related to $T_{h}^{\rm hh}(k;>M_{\rm th},>M'_{\rm th})$ as
\begin{align}
  \left.\parfrac{P_{\rm gg}^{2h}(k)}{\deltab}\right|_{\rm G} 
  &=\frac{1}{\bar{n}_g^2} \iint \mathrm{d}M \mathrm{d}M'  \frac{\mathrm{d}\tilde{U}_g(k;M')}{\mathrm{d}M'} \left[ \frac{\mathrm{d}\langle \tilde{U}_g (k;M)\rangle}{\mathrm{d}M} 
  \left.\parfrac{\mathscr{P}_{\rm hh}(k;>M,>M')}{\deltab}\right|_{{\rm G}, M, M'} \right. \nonumber\\
  &~~~~~ \left. + \frac{2}{3} {\mathscr{P}_{\rm hh}(k;>M,>M')}
\frac{\partial}{\partial \ln k} \frac{\mathrm{d}\langle \tilde{U}_g(k;M)\rangle}{\mathrm{d}M}
    \right] 
  - 2b_g^L P_{\rm gg}^{2h}(k) \nonumber \\
  &\simeq \frac{1}{\bar{n}_g^2} \iint \mathrm{d}M \mathrm{d}M' \left\{ \frac{\mathrm{d}\langle \tilde{U}_g (k;M)\rangle}{\mathrm{d}M} \frac{\mathrm{d}\tilde{U}_g(k;M')}{\mathrm{d}M'} \right. 
  n(>M) n(>M') 
  \left[ 2b_{\mathrm{h}, 1}^L(>M') P_{\rm hh}(k;>M,M') \right. \nonumber \\
  &~~~~~ + \left.\left. \left(\frac{26}{21}\right) T_{h}^{\rm hh}(k;>M,>M') 
   P_{\rm hh}(k;>M,>M') + \frac{2}{3} {P_{\rm hh}(k;>M,>M')}
\frac{\partial}{\partial \ln k} \frac{\mathrm{d} \ln \langle \tilde{U}_g(k;M)\rangle}{\mathrm{d}M}
     \right] \right\}  \nonumber \\
  &~~~~~ - 2b_g^L P_{\rm gg}^{2h}(k). 
  \label{eq:G^gg2h}
\end{align}
Using Eqs.~(\ref{eq:G^gg1h}) and (\ref{eq:G^gg2h}), we can compute the SU growth response for the galaxy-auto power spectrum as Eq.~(\ref{eq:G^gg}).

\twocolumngrid
\bibliography{lssref-short}

\end{document}